\documentclass[manuscript, screen,noacm,oneside]{acmart}

\usepackage{needspace} 
\usepackage{array}
\usepackage{lipsum}
\usepackage{lscape}
\usepackage{amsmath}
\usepackage{capt-of}
\usepackage{caption}
\usepackage{setspace}
\usepackage{verbatim}

\usepackage{soul} \usepackage{graphicx}
\graphicspath{{figures/}}
\DeclareGraphicsExtensions{.jpg,.png}
\usepackage{etoolbox}
\usepackage[caption=false,font=footnotesize]{subfig}

\usepackage{enumitem}
\usepackage{booktabs}

\usepackage{multirow,makecell}

\usepackage{framed}
\usepackage{float}

\usepackage{listings}
\usepackage{rotating}
\usepackage{adjustbox}
\usepackage{tabularx}
\usepackage[framemethod=default]{mdframed}
\usepackage{tikz}
\usepackage{pgf-umlsd}
\usepackage{pgfplots, pgfplotstable}
\usetikzlibrary{arrows,shadows,patterns}

\usepackage[T1]{fontenc}
\usepackage{lmodern}

\usepackage{pgfplotstable}
\usepackage{filecontents}

\usepackage{url}

\newcolumntype{B}[2]{>{\adjustbox{angle=#1,lap=\width-(#2)}\bgroup}l<{\egroup}}

\newcolumntype{Z}[2]{>{\adjustbox{angle=#1,lap=\width-(#2)}\bgroup}l<{\egroup}}

\newcolumntype{L}[1]{>{\raggedright\let\newline\\\arraybackslash\hspace{0pt}}m{#1}}
\newcolumntype{C}[1]{>{\centering\let\newline\\\arraybackslash\hspace{0pt}}m{#1}}
\newcolumntype{R}[1]{>{\raggedleft\let\newline\\\arraybackslash\hspace{0pt}}m{#1}}

\newcommand{\full}{\ $\bullet$}
\newcommand{\prt}{\ $\circ$}

\newcommand{\headrowx}[2]{\multicolumn{1}{l}{\adjustbox{angle=#2,lap=\width-0.5em}{\small #1}}}

\definecolor{semi-light-gray}{gray}{0.7}
\definecolor{light-gray}{gray}{0.8}

\newcolumntype{Y}{>{\centering\arraybackslash \hsize=0.9\hsize}X} \newcolumntype{Z}{>{\centering\arraybackslash \hsize=.5\hsize}X}

\newcommand{\idpspassoc}{IdP-SP Association Model}
\newcommand{\assertauth}{IdP: User Identity Conveyance Method}
\newcommand{\assertchan}{SP: User Identity Verification Method}
\newcommand{\machinetype}{User to IdP Authentication Type}
\newcommand{\devauthtoken}{Multi-Device Usage Model}

\newcommand{\portableid}{Portable-Identity-Across-IdP}
\newcommand{\nodevicepair}{No-Device-Setup}
\newcommand{\nohwtoken}{No-Hardware-Token-Required}
\newcommand{\restempoutage}{Resilient-to-Temporary-Service-Outage}

\newcommand{\noidpvetting}{IdP-Vetting-Not-Needed}
\newcommand{\nospsponsor}{SP-Registration-Not-Needed}
\newcommand{\spusersecret}{No-SP-Stored-User-Secret}

\newcommand{\deviceboundsecret}{Resilient-to-Leaks}
\newcommand{\clientleak}{Resilient-to-Client-Leaks}
\newcommand{\spleak}{Resilient-to-SP-Leaks}
\newcommand{\remoteleak}{Resilient-to-Third-Party-Leaks}
\newcommand{\signalsloa}{Signals-Assurance-Level-to-SPs}
\newcommand{\authidps}{SPs-Can-Filter-IdPs}
\newcommand{\notpimpersonation}{No-Impersonation-by-Third-Party}

\newcommand{\privatefromidp}{Private-Browsing}
\newcommand{\spunlinkable}{Unlinkable-Across-SPs}
\newcommand{\nodatasharing}{No-Sharing-of-User-Data}

\usepackage[ruled]{algorithm2e} 
\SetAlFnt{\small}
\SetAlCapFnt{\small}
\SetAlCapNameFnt{\small}
\SetAlCapHSkip{0pt}
\IncMargin{-\parindent}

\setcopyright{none}

\begin{document}
\settopmatter{printacmref=false}
\fancyfoot{}

\title{Comparative Analysis and Framework Evaluating Web Single Sign-On Systems}

\author{Furkan Alaca}
\affiliation{\institution{School of Computing, Queen's University, furkan.alaca@queensu.ca}
  \city{Kingston} 
  \state{ON} 
}
\email{furkan.alaca@queensu.ca}

\author{Paul C. van Oorschot}
\affiliation{\institution{School of Computer Science, Carleton University, paulv@scs.carleton.ca}
  \city{Ottawa} 
  \state{ON} 
}
\email{paulv@scs.carleton.ca}

\makeatletter
\let\@authorsaddresses\@empty
\makeatother

\begin{abstract}
We perform a comprehensive analysis and comparison of 14 web single sign-on (SSO) systems proposed and/or deployed over the last decade, including federated identity and credential/password management schemes. We identify common design properties and use them to develop a taxonomy for SSO schemes, highlighting the associated trade-offs in benefits (positive attributes) offered. We develop a framework to evaluate the schemes, in which we identify 14 security, usability, deployability, and privacy benefits. We also discuss how differences in priorities between users, service providers (SPs), and identity providers (IdPs) impact the design and deployment of SSO schemes.
\end{abstract}

\maketitle
\thispagestyle{empty} % https://github.com/borisveytsman/acmart/issues/97

\section{Introduction} \label{sec:introduction}

Password-based authentication has numerous documented drawbacks. Despite the many password alternatives, no scheme has thus far offered improved security without trading off deployability or usability \cite{bonneau2012quest}. Single Sign-On (SSO) systems have the potential to strengthen web authentication while largely retaining the benefits of conventional password-based authentication. We analyze in depth the state-of-the art of web SSO through the lens of usability, deployability, security, and privacy, to construct a taxonomy and evaluation framework based on different SSO architecture design properties and their associated benefits and drawbacks.

SSO is an umbrella term for schemes that allow users to rely on a single master credential to access a multitude of online accounts. We categorize SSO schemes across two broad categories: federated identity systems (FIS) and credential managers (CM). FIS establishes a means of communicating user identity across administrative domains; this allows users to authenticate to an Identity Provider (IdP) that can communicate assurances of user identity to Service Providers (SPs; often also referred to as Relying Parties, or RPs), thereby granting users access to SP services without needing to re-authenticate. CM-based SSO stores SP-specific credentials (e.g., passwords or cryptographic keys) and automatically uses them to authenticate to SPs on behalf of users; CMs are typically protected by a single master credential such as a password or a hardware token containing a cryptographic key (e.g., smart card). We classify both FIS- and CM-based SSO schemes into more granular subcategories, while identifying benefits and drawbacks associated with different approaches. While SSO has long been used in enterprise networks to enable users to access network services and applications with a single set of credentials (e.g., Kerberos~\cite{kerberosv5}), we focus specifically on SSO designed for web authentication.

Our primary contributions are summarized as follows:\footnote{An earlier version of this work was made available as a technical report in April 2018 \cite{arxivpreprint}.}
\begin{enumerate}
\item While prior work has explored various aspects of SSO security, this is to our knowledge the first to perform a comprehensive analysis and comparison of a broad range of SSO systems proposed and/or deployed within the last decade, including two hardware-based SSO schemes that are undergoing large-scale deployment, FIDO UAF~\cite{fido_uaf} and Mobile Connect~\cite{mobileconnect}.
\item We identify different design properties of SSO schemes and develop a taxonomy built on a categorization scheme across the design properties. We also develop a corresponding evaluation framework that highlights benefits and drawbacks of SSO schemes based on their design properties, and analyze a representative subset of 14 SSO schemes under this framework.
\item We identify trade-offs between different design goals, by identifying how various SSO schemes can be augmented with existing techniques to achieve specific benefits while forgoing others. Such trade-offs allow SSO schemes to be tailored to different needs and scenarios.
\end{enumerate}

The remainder of this paper is organized as follows. Section~\ref{sec:related_work} discusses background and related work on SSO. Section~\ref{sec:sso_protocols} provides an overview of the SSO protocols evaluated in this paper. Section~\ref{sec:sso_properties} presents the new classification and evaluation framework, and Section~\ref{sec:sso_compare} provides a detailed evaluation of each scheme. Section~\ref{sec:discussion} discusses insights from the evaluation, and Section~\ref{sec:conclusion} concludes.

\section{Background and Related Work} \label{sec:related_work}
Bonneau et al. \cite{bonneau2012quest} evaluate 35 authentication schemes, including a number of SSO systems, based on a usability, deployability, and security (UDS) framework. Among all categories of schemes considered, FIS (e.g., OpenID) and CM (e.g., password managers, FIDO UAF) schemes retained the most usability and deployability benefits of passwords, while improving security. The usability benefits of SSO stem from the reduced user memory burden from having to remember fewer passwords---however, while password managers offer the benefit of working with all password-based websites, FIS schemes have the drawback of having much more limited SP support. The security benefit of SSO in general hinges on the assumption that users will be able to choose and remember a stronger master password (since they will have fewer passwords to remember); additionally, the security benefits offered by CM-based schemes rely on users picking (or randomly generating) unique passwords across SP accounts, to provide resilience against online guessing attacks and to limit the damage caused by any single compromised password to a single corresponding SP. Our analysis of SSO schemes is applicable with or without these assumptions, and our taxonomy and evaluation framework are complementary to the UDS framework in that we identify and analyze additional usability, deployability, security, and privacy benefits specifically relevant to SSO systems, independent of the user authentication mechanism used.

NIST's Digital Identity Guidelines~\cite{nist_digidentity_2017,nist_digidentity_fed_2017}, as revised in 2017, provide technical and procedural guidelines for US government agencies implementing several FIS models, which are part of our analysis in Section~\ref{sec:idpspassoc}. Pashalidis and Mitchell \cite{pashalidis2003taxonomy} categorize web SSO broadly into \textit{pseudo-SSOs}, \textit{true SSOs}, \textit{proxy-based SSOs}, and \textit{local SSOs}; the first two categories correspond loosely to what we call FIS and CM schemes, and the latter two correspond loosely to the \textbf{T1} and \textbf{T2} categories we define in Section~\ref{sec:machinetype}. Our taxonomy and evaluation framework, developed as a result of an analysis of newer schemes that have been proposed and deployed since the analysis of Pashalidis and Mitchell, provides a more granular means by which current schemes can be evaluated and compared.

Below we summarize existing research related to schemes evaluated in our analysis, leaving our technical overviews  of individual schemes to Section~\ref{sec:sso_protocols}.

\textbf{SSO vulnerabilities.} Whereas our taxonomy and evaluation framework for SSO schemes is based on high-level design properties, in practice SSO security also depends on the secure design and implementation of the underlying protocols. Many SSO vulnerabilities arise due to widespread implementation errors (an analogy can be drawn to the widespread practice of websites using password authentication that insecurely store passwords). Sun et al.~\cite{sun2012oauth} analyze 96 popular OAuth 2.0 based SPs using Facebook SSO, and find that the majority are vulnerable to at least one serious implementation-related vulnerability such as access token theft via a cross-site scripting (XSS) attack. Chen et al.~\cite{chen2014oauthmobile} analyze over 600 mobile applications using OAuth 2.0 for SSO authentication, and find that about 60\% were vulnerable to exploit due to incorrect implementation. Fett et al. \cite{fett2014browserid} develop a formal analysis framework for web SSO implementations, and apply it to BrowserID (the protocol underlying Mozilla Persona) to find a number of server-side IdP implementation-related vulnerabilities; they also \cite{fett_oidc_2017} conduct a formal analysis of OpenID Connect to discover new classes of attacks and outline implementation guidelines for corresponding defenses. Mainka et al.~\cite{mainka_donottrust_2016} discover implementation-related SP vulnerabilities that allow attackers to impersonate users by deploying malicious IdPs to masquerade as other legitimate IdPs (these attacks target server-side SP implementations, and are distinct from phishing attacks that target users).

Some SP vulnerabilities can be traced back to inadequate documentation in protocol specifications or SDKs. Sun et al.~\cite{sun2012openidsecurity} conducted a formal analysis of OpenID 2.0 and found that protection against session tampering required the implementation of safeguards not discussed in the specifications; an empirical study revealed that many OpenID 2.0 RPs were vulnerable to cross-site request forgery (CSRF) attacks since they did not have the safeguards in place. Wang et al.~\cite{wang2013explicatingsdks} formally analyzed OAuth 2.0 SDKs provided by Microsoft and Facebook (two major IdPs) to determine whether SP developers can securely build SSO into their applications by following only explicitly-stated assumptions in the SDKs; it was found that many apps built with these SDKs were vulnerable to major exploits due to violations of unstated assumptions by IdPs (e.g., the expected sequence of API calls made by an SP). 

Implementation-related vulnerabilities can also impact password managers (but potentially to a lesser degree than FIS---IdP-side vulnerabilities may be more likely to be promptly fixed than SP-side vulnerabilities \cite{zhou2014ssoscan}). For example, flawed password auto-fill policies may result in password compromise \cite{silver2014pwdmanagers} by, e.g., filling in the user's password on a page retrieved via a non-HTTPS connection. Web-based password managers that operate by locally injecting JavaScript into websites via a browser extension or bookmarklet can be vulnerable to various XSS or CSRF attacks \cite{li2014emperor}.

\textbf{Automated SP vulnerability scanners.} Although implementation-related vulnerabilities are widespread, automated vulnerability scanners can help SPs secure their applications by remotely scanning for common implementation errors. Zhou et al.~\cite{zhou2014ssoscan} develop an automatic vulnerability checker to test for five different vulnerabilities; they scanned over 1660 sites that use Facebook SSO, finding that about 20\% were susceptible to at least one vulnerability. Mainka et al.~\cite{mainka_oidc_2017} analyze common implementation-based vulnerabilities, categorize them (e.g., one category is replay attacks allowing attackers to impersonate users by sending expired access tokens to SPs that fail to check the nonce or expiry parameters), and develop an automated vulnerability assessment tool to scan OpenID Connect SPs for common vulnerabilities; they also discovered protocol-level vulnerabilities in OpenID Connect.

\textbf{Hardening SSO protocols.} A common threat to virtually all current web authentication schemes, including SSO, is that of session hijacking. Typically, successful user authentication to a website is followed by the provision of an HTTP session cookie allowing users to browse across different pages of the website without re-authenticating. Session cookie theft (e.g., via XSS attacks) allows attackers to bypass user authentication by hijacking an existing authenticated user's session. Token Binding~\cite{ietf_tokenbinding} (formerly TLS ChannelID \cite{ietf_channelid} and Origin-Bound Certificates \cite{dietz_obc_2012}) allows cryptographically binding tokens (e.g., session cookies, access tokens sent from IdPs to SPs via the user's browser) to browsers using client-side dynamically generated TLS certificates \cite{grosse_authentication_2013}. Token binding can be applied across different SSO protocols to defend against various token theft attacks (such as session cookie theft or identity assertion reuse), e.g., Dietz et al. \cite{dietz2014persona} implement token binding for Mozilla Persona, and FIDO UAF~\cite{fido_tokenbinding} supports it as an optional feature (for backwards-compatibility with platforms that do not yet support token binding).

\textbf{SSO adoption.} Sun et al. investigate SSO adoption barriers both from users'  \cite{sun2011usersrefuse} and SPs' \cite{sun2010billion} perspectives. Their user study reveals various reasons behind users' hesitance to adopt SSO, such as inaccurate user mental models (e.g., incorrectly believing that they were sharing their password with the RP) and concerns about privacy and potential phishing attacks. In contrast, a user study by Stobert~\cite{stobert2014pwlifecycle} indicates that users are much more willing to adopt password managers built into the OS or web browser---this motivates our analysis of CM-based schemes such as Firefox Sync and FIDO UAF\footnote{FIDO UAF is currently built into a number of major platforms that advertise it under their own branding, e.g., ``Windows Hello'' on Windows 10 and ``Samsung Pass'' on Samsung smartphones. Related to this are standards such as the W3C Web Authentication API~\cite{w3cwebauthn} that support the integration of FIDO UAF authenticators into browsers and operating systems.}. McCarney~\cite{mccarney2013thesis} discusses other classes of password managers, e.g., managers that use a master password to generate SP-specific passwords---such schemes have not received much interest, due to practical issues such as the need to reset all SP passwords when the master password is changed.

\section{Overview of SSO Protocols} \label{sec:sso_protocols}

To give technical context for our comparative analysis, we give a simplified overview of the schemes analyzed. Pointers to in-depth overviews are referred to by citations herein.

\subsection{Credential Managers}

As cited in Section~\ref{sec:related_work}, CM-based schemes seem to gain wider acceptance from users when they are already built into their browser or OS. This motivates our analysis of Firefox Sync~\cite{ffsync} (password-based CM) and FIDO UAF~\cite{fido_uaf} (public-key based CM). We also analyze Impostor~\cite{pashalidis_impostor_2004}, since its properties help demonstrate our taxonomy. 

\subsubsection{Firefox Sync} \label{sec:sync_overview}
The Firefox password manager saves user-created passwords (an updated interface will offer a random password generator \cite{moz_lockbox_faq}) when typed into a website, and offers to automatically enter them on subsequent visits. Users may also manually view their saved passwords through a graphical interface. The password database is stored locally in a file, which the Firefox Sync~\cite{ffsync} protocol synchronizes across users' devices. Below, we give a simplified overview of two versions of this protocol. 

\textbf{Firefox Sync 1.5} locally generates a symmetric key to encrypt the user password database. The key is stored locally, and the encrypted password database is uploaded to the user's Sync account, accessed with a user name and password that is only typed in once per device during the initial Sync setup \cite{warner_ffsync}. After setting up Sync on one device, the symmetric key can be transferred to another user device over a secure connection initiated via a ``pairing'' process that first establishes a shared session key. The pairing process uses J-PAKE \cite{hao2010jpake} to display a short code on one device, which the user types into the second device. User devices remain synchronized by downloading the latest copy of the encrypted password database from the Sync server, and uploading any updated contents if necessary. Sync 1.5 was deprecated, since many users did not understand the purpose of the pairing process and expected to access their password database from any device using only their Sync password. Users were generally unaware of the existence of the encryption key and of the option to print it out for backup purposes, and consequently many users lost access to their passwords if they only owned a single device that they replaced without first backing up their key.

\textbf{Firefox Sync 2.0} \cite{ffonepw} locally derives two symmetric keys from the user's Sync password using PBKDF2~\cite{pbkdf2} and HKDF~\cite{hkdf}. One key is used for authenticating to the user's Sync account (thereby not revealing the Sync password to the server), and the second key is used to encrypt the password database before uploading it to the server. This eliminates the need for device pairing; setting up a new device only requires the user to enter their password during Sync setup. Although iterated hashing is used both on the client side (to generate the authentication key) and on the server side (for storing a hashed version of the authentication key) to slow down offline attacks, the strength of the encryption key is still dependent on the user-chosen Sync password (in contrast to Sync 1.5, where the encryption key is randomly generated).

By default, both versions of Firefox Sync store passwords unencrypted on the client system. Users may optionally select an offline master password (separate from the synchronization password), from which a key is derived to encrypt the password database on-disk. Chrome Sync~\cite{chromesync}, which employs a synchronization protocol similar to that of Firefox Sync 2.0, uses the OS's built-in credentials manager (which typically encrypts its contents with users' login passwords) to store passwords client-side. Many commercial password managers also employ similar synchronization protocols, but are excluded from our analysis both due to their high degree of technical similarity to the browser-based password managers analyzed herein and since average users prefer options that are built into their browser or OS~\cite{stobert2014pwlifecycle}.

\subsubsection{FIDO UAF}

FIDO UAF~\cite{fido_uaf} (Universal Authentication Framework) uses client-side UAF authenticator devices to authenticate users to SPs via public-key cryptography. UAF authenticators are typically hardware modules built into end-user devices, but they may also be software-based. When users register a UAF device with an SP account (each device must be individually registered with each SP), the UAF client generates an SP-specific key pair (private keys are stored on-device; corresponding public key certificates, which are either self-signed or signed using an attestation key discussed below, are sent to SPs). Local authentication, such as a PIN or biometric, is used to ``unlock'' the UAF authenticator and initiate the cryptographic challenge-response protocol to authenticate users to SPs.

Hardware-based UAF authenticators with secure key storage capabilities can be certified by the FIDO Alliance and provisioned with a signed attestation certificate containing metadata about the device (e.g., device manufacturer, method of local user authentication used). SPs can validate the certificate for a higher degree of user identity assurance, since hardware-based UAF clients can protect against key theft to a much higher degree than software-based UAF clients. Attestation certificates can be revoked in the event that vulnerabilities are found (e.g., which may allow the extraction of private keys from the device) in a hardware UAF client, to aid SPs in phasing out support for vulnerable hardware.

FIDO U2F (Universal Second Factor)~\cite{fido_u2f} is a related open standard for two-factor authentication, of considerable industrial interest, but by definition is designed to be used as a second factor, whereas our study of SSO schemes is by definition focused on stand-alone schemes (despite the fact that a U2F token, if used stand-alone, would fit the definition of a CM scheme). Usability of several U2F devices has also been studied \cite{reynolds2018tale,das2018johnny,lang2016security}.

\subsubsection{Impostor} \label{sec:ov_shibboleth}
Impostor~\cite{pashalidis_impostor_2004} is a proxy-based CM that stores users' passwords on a remote IdP. Passwords are automatically submitted to SPs that the user visits through the IdP proxy. Its design goal was to provide a means of using an untrusted machine without exposing any long-term secret to it. Therefore, the authors suggest that the user-to-IdP authentication be done through a challenge-response based scheme, e.g., a hardware one-time-password (OTP) token (which we assume in our analysis).

\subsection{Federated Identity Systems}
An identity federation consists of one or more member IdPs and one or more member SPs such that users can authenticate to member SPs through member IdPs. The authentication process consists of a user-to-IdP authentication stage, followed by an IdP-to-SP identity assertion to convey the user's identity information (Section~\ref{sec:assertauth} outlines different methods for doing so) and assure the SP that the user has been authenticated by the IdP. FIS protocols can have proprietary or open specifications---all protocols discussed herein have open specifications. Some schemes that we evaluate are protocols (such as OpenID 2.0, OAuth 2.0, and OpenID Connect), and others are specific implementations of protocols, such as Mobile Connect (a proprietary implementation of OpenID Connect) and Shibboleth (a free implementation of SAML2). The set of schemes we evaluate were selected to best highlight the features of our taxonomy and evaluation framework.

Different organizations using the same protocol are not necessarily part of the same federation. Moreover, both IdPs and SPs may support SSO authentication via multiple protocols and be a member of multiple federations. Section~\ref{sec:idpspassoc} discusses different types of federations based on how IdP-SP associations are established. 

\subsubsection{Shibboleth} \label{sec:shibboleth}
Shibboleth~\cite{shibboleth} is a popular open-source FIS implementation based on SAML~\cite{oasis}, a highly flexible and extensible XML-based standard for exchanging identity information between federation members. SAML's high flexibility requires many protocol message format details to be agreed upon between federation members, thereby reducing deployment scalability \cite{sun2013thesis}. A high-level overview of the protocol is as follows:
\begin{enumerate}
\item The user visits an SP, and clicks the login link to initiate the authentication process.
\item A discovery process (see Section~\ref{sec:idpspassoc}) takes place to discover the user's IdP, e.g., by presenting a list of supported IdPs to the user.
\item The user is redirected to their IdP. If an authenticated session does not already exist with the IdP, the user logs in.
\item The IdP generates an authentication response and attaches it to an HTTP POST request when redirecting the user back to the SP (while the authenticity and integrity of the authentication response is protected by TLS, it may also be signed by the IdP and encrypted with the SP's key as an additional layer of protection \cite{shib_flows}).
\item The SP receives and validates the authentication response, and creates an authenticated session for the user.
\end{enumerate}
\subsubsection{OpenID 2.0} \label{sec:ov_oid2}
The OpenID 1.x \cite{recordon2006openid} and 2.0 \cite{openid2} family of specifications were designed to be a much simpler (but also less flexible) FIS scheme compared to SAML. The simpler protocol message format enables OpenID-based IdPs and SPs to communicate without requiring prior agreement on a large set of protocol parameters as was the case with SAML. The design philosophy of OpenID was to allow any domain owner to set up an IdP (users can therefore set up their own IdP if desired) and provide services to any SP without prior coordination. The OpenID user ID format is a URL (or XRI~\cite{oasis_xri}) of a user profile page; the page contains metadata that points SPs to the IdP URL to which the user should be redirected for authentication. The high-level protocol flow is similar to that of Shibboleth as described above, but the IdP discovery step consists of retrieving the metadata from the user's supplied profile URL. OpenID SPs experimented \cite{sun2011usersrefuse,sachs_openid_2008} with various user interface designs for obtaining users' OpenID identifiers---generally, users would select their IdP from a list of logos of the most popular IdPs, and subsequently type in their user name (the SP could then determine the correct IdP URL corresponding to the user name); more advanced users could pick the option to manually enter their own URL instead (e.g., if they host their own IdP or use a lesser-known IdP).

\subsubsection{OAuth 2.0} \label{sec:ov_oauth2}

OAuth 1.0 \cite{oauth1} and 2.0 \cite{oauth2} enable users to authorize web applications to retrieve resources (e.g., photos or documents) from or perform actions (e.g., upload a new document) on a user account on a resource server without revealing their account password to web applications. However, OAuth can also be used as a FIS if the role of the resource server is to store identity information.  A high-level overview of OAuth 2.0 as an authorization protocol is as follows:
\begin{enumerate}
\item The user visits an SP and initiates the authentication process, typically by clicking on a button from among a list of supported IdPs.
\item The user is redirected to the IdP for authentication, and authenticates to the IdP.
\item The IdP presents the user with a list of resources that the SP has requested permission to access. At the minimum, the SP requires access to the user's identity information. However, the SP may request additional permissions (e.g., accessing the user's contact list or permission to make social media posts on the user's behalf).
\item If the user approves the permissions requested by the SP, the IdP generates an authorization code and encodes it as a URI parameter when redirecting the user back to the SP.
\item The SP submits the authorization code to the IdP in exchange for an access token. Access tokens may be short-lived, when used by an SP to only verify a user's identity and establish an authenticated session, but may also be long-lived if SPs request persistent access to perform actions on the user's IdP account (e.g., periodically obtain an updated copy of the user's social media contacts) even while the user is not logged into the SP.
\item The SP queries the IdP to obtain a user identifier (e.g., e-mail address) that is associated with the access token; the SP then creates an authenticated session (i.e., initializes the server-side session, and provides a corresponding session cookie to the browser) for the account corresponding to the user identifier.
\end{enumerate}

Unlike OpenID, OAuth requires SPs to register out-of-band with IdPs they wish to support. The registration typically requires the SP administrator to submit an online form on the IdP website and establish a shared secret (typically a randomly-generated string) for the SP to include when making API calls to the IdP. The primary purpose of the shared secret is to prevent unauthorized use of access tokens (in case of token theft), which provide access to user information and resources (albeit limited by the permissions granted by the user in step 3 above) stored by the IdP.  OAuth 2.0 relies on TLS to protect the secrecy of the shared secret.

\subsubsection{OpenID Connect} \label{sec:ov_oidc}
OpenID Connect \cite{openidconnect} defines a standardized mechanism for exchanging identity information over OAuth 2.0. The high-level protocol overview is as described above for OAuth 2.0, except that to identify the user in the final step, the SP requests an ID token (in a standard JSON Web Token~\cite{ietf_jwt} format) from the IdP. The more strictly-defined protocol message format, compared to OAuth 2.0, facilitates building code libraries for SPs that can interoperate with multiple IdPs. OpenID Connect also includes a number of optional (but rarely used) features such as \textit{dynamic client registration}, which allows SPs to register with IdPs automatically instead of via an out-of-band process; this is intended to allow OpenID Connect to operate more similarly to OpenID 2.0, i.e., to allow any SP to request user authentication services from any IdP, without any prior coordination. OpenID Connect also defines a framework for establishing federated relationships; this allows co-operating IdPs to form a federation, and allows SPs to register with the federation (instead of with each individual IdP) to gain access to the services of all member IdPs.

\subsubsection{Mobile Connect} \label{sec:ov_mobileconnect}
Mobile Connect \cite{mobileconnect} is an OpenID Connect federation operated by the GSM Association; its IdPs consist of mobile network operators (MNOs) worldwide, and users authenticate to IdPs using their mobile phones as hardware authenticator tokens. Mobile Connect redefines several optional API parameters as mandatory \cite{mobileconnect_openid}, such as the \textsf{nonce} and \textsf{state} parameters used for binding an access token to an HTTP session; the \textsf{acr\_values} parameter is used by SPs to communicate a required Level of Assurance (LoA) to IdPs for user authentication on a 4-point scale, as defined by ISO/IEC and ITU-T~\cite{iso_iec_29115,itu-t_x1254}:
\begin{itemize}
\item LoA1: Minimal confidence that user's identity is consistent across multiple authenticated sessions (e.g., using a device's MAC address). Not applicable to Mobile Connect.
\item LoA2: Some confidence in user's asserted identity. Requires at least one authentication factor; typically implemented by Mobile Connect IdPs by sending an SMS one-time password (OTP) to their mobile phone (to verify possession), which the user types into the device (e.g., computer) they are authenticating from.
\item LoA3: High confidence in user's asserted identity. Requires at least two authentication factors; typically implemented by Mobile Connect IdPs by requiring a user-to-device authentication mechanism (on the mobile device) such as a 4-digit PIN or a biometric before displaying the OTP that the user needs to type in.
\item LoA4: Similar to LoA3, but requires in-person identity proofing (i.e., the user's online identity is associated with a real-world individual). Not currently supported by Mobile Connect.
\end{itemize}

\subsubsection{Mozilla Persona} \label{sec:ov_persona}
Mozilla Persona \cite{hanson2011federated} was\footnote{Due to limited adoption, Mozilla discontinued internal development of the project in 2014 \cite{persona_transition}. Nevertheless, it remains an interesting protocol to study due to its unique approach and the interest it received in the security community.} designed to enable users to tie their online identities to their e-mail addresses, and for e-mail providers to fill the role of IdPs for their users. Through public-key cryptography, Persona IdPs delegate to users' browsers the responsibility of generating and sending identity assertions to SPs; this has the privacy benefit that IdPs do not learn of the SPs that users visit. Similarly to OpenID 2.0, automatic discovery (i.e., a protocol mechanism allowing for SPs to request service from IdPs without any prerequisite human intervention such as out-of-band key establishment) is supported, allowing any Persona SP to request user authentication services from any Persona IdP. A high-level protocol overview is as follows:
\begin{enumerate}
\item The user visits an SP website and clicks on a ``Log in with Persona'' button, which pops up a new window in which the user enters their e-mail address.
\item The browser (using client-side JavaScript) determines whether the user's e-mail provider supports Persona by checking for the presence of a \textsf{browserid} file accessible from the e-mail provider's domain, at the location \textsf{https://idp.domain/.well-known/browserid}. This file includes the IdP's public key used for signing certificates, and the URL at which the IdP's users should be redirected for authentication.
\item The user is redirected to the URL specified by the \textsf{browserid} file, and authenticates to their IdP.
\item The browser (using client-side JavaScript) generates a public-private key pair and sends the public key to the IdP. The IdP returns a signed \emph{user certificate}, containing the browser's public key.
\item The browser generates and signs an identity assertion for the SP that the user wishes to authenticate to. The signed assertion is sent to the SP, along with the user certificate from the previous step.
\item The SP verifies the IdP's signature (the IdP's public key can be obtained from the \textsf{browerid} file discussed in step 2) on the user certificate, and subsequently the browser's signature on the identity assertion.
\item The SP initiates an authenticated session for the user.
\end{enumerate}

Due to limited adoption of Persona by e-mail providers, Mozilla introduced a \emph{fallback IdP} to authenticate users whose e-mail providers did not support Persona. Users could create a Persona account on the Mozilla fallback IdP, by using their e-mail address as their user name. Upon account creation, the fallback IdP e-mails the user a verification URL, which the user clicks to verify that they control the e-mail address. Creating the Persona account requires users to select a password, to be used for authentication in step 3 above, when the user is forwarded from an SP to the Mozilla fallback IdP URL for authentication.

For some e-mail providers that support OAuth 2.0, the Mozilla fallback IdP acted as an OAuth 2.0 bridge: instead of the user having to create a password for their Persona account and verifying possession of their e-mail address through a verification URL, users were instead redirected to their e-mail provider to authenticate via OAuth 2.0. In effect, the Mozilla fallback IdP acts as an SP to request an OAuth access token (see Section~\ref{sec:ov_oauth2}) from the user's e-mail provider. The fallback IdP then uses the access token to obtain the user's e-mail address from the e-mail provider, and compares it with the e-mail address provided by the user to verify the user's possession of the e-mail address. The benefit of this approach (compared to the default behaviour of the fallback IdP as described above) is that the user does not have to create a new password for use with Persona.

\subsubsection{SecureKey Concierge} \label{sec:ov_securekey}
SecureKey Concierge \cite{sktrustfw} is a privacy-focused SSO system that enables IdPs to provide user authentication to SPs through an intermediary service that performs triple blinding: SPs are blinded from users' selected IdPs, IdPs are blinded from the SPs that users access, and SecureKey is blinded from any personally-identifying user information. SecureKey is used by various online services offered by the Government of Canada, and the approved list of IdPs currently consists of Canadian banking institutions. A high-level protocol overview is as follows:
\begin{enumerate}
\item The user visits an SP and initiates the authentication process by clicking on a link to authenticate through SecureKey Concierge. 
\item The user is prompted to choose from a list of IdPs approved by SecureKey.
\item SecureKey forwards the user to their selected IdP for authentication.
\item Upon successfully authenticating the user, the IdP generates a meaningless-but-unique identifier (MBUN) for the user. The user is then redirected back to SecureKey, along with the MBUN. The MBUN blinds SecureKey from users' real-world identities.
\item SecureKey internally maps the MBUN to an internal Persistent Anonymous Identifier (iPAI). The iPAI is used to generate an RP-specific (i.e., SP-specific) Persistent Anonymous Identifier (rpPAI) to forward back to the SP. Generating a unique rpPAI for each SP prevents colluding SPs from correlating the user's identity across accounts.
\item SecureKey redirects the user back to the SP, along with the rpPAI.
\item The SP initiates an authenticated session for the user.
\end{enumerate}

The protocol flow as described above includes two authentication flows chained together:
\begin{enumerate}
\item SecureKey acts as an IdP when interacting (over either Shibboleth or OpenID Connect) with SPs. The SP forwards users to SecureKey for authentication, and when the process is completed the SP only receives an rpPAI in an authentication response message signed by SecureKey.
\item SecureKey acts as an SP when interacting (over Shibboleth) with the user's selected IdP: SecureKey forwards users to their selected IdP for authentication, and receives a signed authentication response containing the MBUN.
\end{enumerate}

\subsubsection{SAW} \label{sec:saw}
SAW~\cite{vanderhorst2007saw} leverages e-mail addresses and the existing SMTP e-mail system to build a FIS. The protocol overview is as follows:
\begin{enumerate}
\item The user visits an SP and types in their e-mail address to initiate authentication.
\item The SP sends the user an e-mail containing a verification link to complete the authentication process. The link contains a string that combines a string generated by the SP with a string generated by the user's browser, thereby binding the verification link to the user's session.
\item The user clicks the link, and the SP initiates an authenticated session.
\end{enumerate}

\section{Classification and Evaluation Framework} \label{sec:sso_properties}

To evaluate the SSO schemes discussed in Section~\ref{sec:sso_protocols}, we first define 14 benefits (Section~\ref{sec:benefits}) related to usability, deployability, security, and privacy. We then classify schemes across 5 design properties (defined in Sections~\ref{sec:idpspassoc} through \ref{sec:devauthtoken}), together forming a taxonomy for SSO schemes; each of these sections define categories associated with the respective property, and conclude with an assessment of the benefits that can or cannot be provided by schemes in each category.

\vspace{-1mm}
\subsection{Benefits Provided} \label{sec:benefits}
Below, we define 14 benefits relating to usability, deployability, security, and privacy aspects of SSO systems. 

\vspace{1mm}
\subsubsection{Usability} ~

\textbf{B1: \portableid.} Users can change their IdP without updating their SP accounts (i.e., without having to reconfigure each of their SP accounts to point to their new IdP). This is a usability benefit, as users may wish to change their IdP due to, e.g., changes in policy or pricing from their existing IdP, wanting to benefit from features offered by another IdP, replacing a hardware authenticator, or wanting to host their own IdP. This benefit is partially provided by schemes where the user can only change to another vetted IdP (see B5: \emph{\noidpvetting} benefit) within the same federation. 

\textbf{B2: \nodevicepair.} No software or hardware configuration (e.g., generating or transferring cryptographic keys) is required by the user when authenticating from a new device.

\textbf{B3: \nohwtoken.} Users do not need to carry around a hardware authenticator token. Schemes that restrict users to a single hardware authenticator, such as a mobile phone, lack this benefit. Schemes that allow IdPs to specify their own authentication mechanism (see Section~\ref{sec:machinetype} for more detailed discussion on user-to-IdP authentication) are assumed here to use conventional password-based authentication (unless stated otherwise), since this is the widespread practice. 

\textbf{B4: \restempoutage.} Users can continue to authenticate to SPs even during a temporary outage of a remote server providing an SSO service. As discussed further in this section, the type of SSO scheme determines the remotely-hosted service being relied upon, e.g., an IdP, discovery server, or synchronization server. This is also a deployability benefit, since it reduces the risk for SPs that they will lose user traffic in the event of an IdP outage.

\vspace{2mm}
\needspace{3\baselineskip}
\subsubsection{Deployability} ~

\textbf{B5: \noidpvetting.} IdPs do not need to register with a federation operator, thereby facilitating deployment. This also provides users with a wider range of IdP choices, and even allows them to host their own IdP.

\textbf{B6: \nospsponsor.} SPs do not need to manually register with each individual IdP that they wish to support, thereby facilitating deployment. 

\textbf{B7: \spusersecret.} The SP does not need to store any user secret, thereby facilitating SP implementation and eliminating the possibility of leaking secret credentials in the event of an SP server compromise.

\vspace{2mm}
\subsubsection{Security} ~

The following three properties (B8a, B8b, and B8c) reflect whether the ability of attackers to extract information from three different entities (user device, remote IdP, and SP, respectively) can be leveraged to impersonate the user. We exclude \emph{active} attacks, i.e., involving attackers with persistent and full control over a user device, since this is an extremely challenging threat model for any authentication scheme to defend against. We also exclude session hijacking from our analysis, since SSO does not offer any inherent defense against such attacks---instead, defense against session hijacking can be offered by complementary mechanisms such as token binding (cf. Section~\ref{sec:related_work}).

\textbf{B8a: \clientleak.} Attackers cannot defeat user authentication by extracting (e.g., via malware) data from users' access devices, such as keystrokes or data stored on disk or memory.

\textbf{B8b: \spleak.} Attackers cannot defeat user authentication by extracting (e.g., via server-side software vulnerabilities) user-specific data from the SP, such as passwords or cryptographic authentication keys.\footnote{Extraction of server-specific data, such as TLS keys that facilitate man-in-the-middle attacks against \emph{all} users of the SP, are excluded. SSO does not offer any inherent defense against such attacks, and complementary mechanisms are needed.}

\textbf{B8c: \remoteleak~(e.g., IdP-leaks).} Attackers cannot defeat user authentication by extracting (e.g., via server-side software vulnerabilities) user-specific data from a trusted remote server (e.g., an IdP, bridge IdP, or synchronization server), such as passwords or cryptographic authentication keys.

\textbf{B9: \signalsloa.} This is a security benefit, whereby the IdP can convey to SPs the level of assurance (LoA) that it can provide in the user's identity. LoA is typically dictated based on the strength of the authentication mechanism used. For example, users authenticated with two-factor authentication would have a higher LoA than users authenticated with only a password.

\textbf{B10: \authidps.} SPs can restrict the set of IdPs that they wish to support (via whitelist or blacklist), based on any criteria the SP chooses. For example, based on their needs, SPs may prefer to only support IdPs that offer certain security measures such as \emph{\signalsloa}, IdPs that use two-factor authentication, or IdPs that tie user identity to a real-world attribute such as a telephone number or physical address.

\textbf{B11: \notpimpersonation.} The SSO scheme does not provide any remote server with the ability to impersonate a user. A scheme partially provides this benefit if impersonation attempts can be detected by the user. For example, with conventional password-based authentication with e-mail recovery, a malicious e-mail provider may force a password reset on the user's account to gain access, but this is detectable by users since it results in a denial-of-service.

\vspace{2mm}
\subsubsection{Privacy} ~

\textbf{B12: \privatefromidp.} The IdP has no knowledge of the SPs that its users authenticate to. We also provide this benefit to schemes where the IdP is hosted on a device that is under the user's control.

\textbf{B13: \spunlinkable.} SPs are not provided with any user identifiers that are linkable across different SP accounts. When evaluating schemes for this benefit, we do not consider other mechanisms that colluding SPs may employ (e.g., browser fingerprinting \cite{eckersley2010unique}, or collecting personally-identifiable information from the user such as their e-mail address) to find links between different SP accounts. 

\textbf{B14: \nodatasharing.} Schemes that authorize SPs to access user data from IdPs do not offer this benefit. For example, OAuth 2.0 and OpenID Connect are widely used by major IdPs (e.g., Google and Facebook) to provide SPs with access to account data (e.g., contact lists and demographic information), and provision long-term access tokens that allow SPs to access the data from IdPs even when the user is logged off.

\subsection{\idpspassoc} \label{sec:idpspassoc}
\begin{figure}
\includegraphics[width=0.9\textwidth]{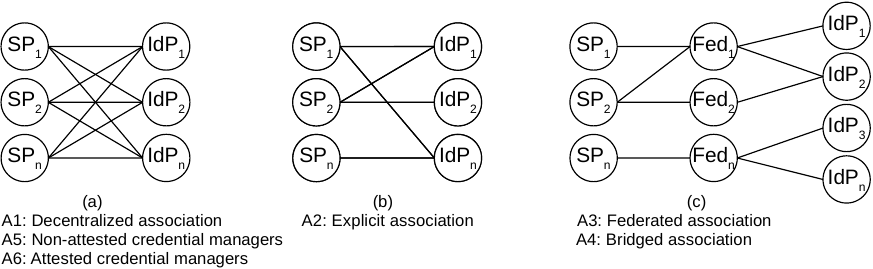}
\caption{IdP-SP association models A1-A6, defined in Section~\ref{sec:idpspassoc}, categorized by IdP-SP relationship graphs. \{A1,A5,A6\} allow users to authenticate to any SP through any IdP; A2 requires IdPs and SPs to have a pre-established relationship, e.g., via a manual registration process; \{A3, A4\} allow users to authenticate to SPs through an IdP only if both the SP and IdP are federation members. \label{fig:trusttax}}

\renewcommand{\arraystretch}{1}
\small
\centering
\resizebox{\linewidth}{!}{
\begin{tabular}{l@{\hskip 0.15in} l@{\hskip 0.15in}l@{\hskip 0.175in} l@{\hskip 8px}l@{\hskip 8px}l@{\hskip 0.3in} L{9cm}}

\headrowx{\idpspassoc}{55} & \headrowx{Association Category}{55} & Example Scheme & \headrowx{NS Auth.: IdP}{55} & \headrowx{NS Auth.: Fed. Operator}{55} & \headrowx{NS Auth.: SP}{55} &
Defining Characteristics
\\\hline

A1 & (a) & OpenID 2.0 & \full &  &  & Any server implementing the protocol can become an IdP or SP.
\\\hline
A2 & (b) & OAuth 2.0 & \full &  &  & SPs establish explicit associations with IdPs, e.g.,  by manual registration to establish shared secrets.
\\\hline
A3 & (c) & Mobile Connect & \full &  &  & A federation operator records which IdPs and SPs belong to this federation; SPs then rely on IdPs in the federation without having explicit relationships with them.
\\\hline
A4 & (c) & SecureKey Concierge &  & \full &  & Similar to A4, except that the federation operator is responsible not only for recording which IdPs and SPs belong to the federation, but also for conveying identity assertions to SPs on behalf of IdPs.
\\\hline
A5 & (a) & Firefox Sync &  &  & \full & Credential management schemes that store and automate the use of SP-specific user credentials.
\\\hline
A6 & (a) & FIDO UAF (Attested) &  &  & \full & Credential management schemes that can provide cryptographic attestation to SPs to provide security assurances, e.g., so the SP can verify the method of user-to-device authentication used.
\\\hline
\end{tabular}}

 \captionof{table}{Summary of IdP-SP association models from Section~\ref{sec:idpspassoc}. Association category (second column) corresponds to the three graphs from Fig.~\ref{fig:trusttax}. For each association model (row), a bullet in one of three columns indicates whether the user identity namespace (NS) authority is the user's IdP, a federation operator, or the SP.\label{table:sso_taxonomy}}
 \vspace{-1mm}
\end{figure}

Figure~\ref{fig:trusttax} illustrates different models by which IdP-SP associations are established. Typically, each IdP is associated with its own namespace. For example, user \emph{John Smith} from IdP\textsubscript{1} and user \emph{John Smith} from IdP\textsubscript{2} are separate identities. Each user-IdP pair is expressed by a unique representation whose format is specified by the SSO protocol, e.g., OpenID 2.0 designates a unique URL for each user under the IdP's domain, and Mozilla Persona uses e-mail addresses under the IdP's domain. An IdP-SP association is characterized by an SP allowing users to identify themselves by their ownership or control over an account or resource under the IdP's namespace. Such an identification process requires a protocol that enables the user to prove their control over the IdP account. We describe six models of IdP-SP association below, and provide a summary in Table~\ref{table:sso_taxonomy}. For each model, we also specify whether the user identity namespace is managed by the user's IdP, a federation operator, or whether each SP manages its own user identity namespace (e.g., as is the case with CM schemes). Models A1-A4 cover FIS schemes, and A5-A6 cover CM schemes.

\textbf{A1: Decentralized association (IdP-managed namespace).} This encompasses ``open'' protocols in which any entity implementing the protocol can become an IdP or an SP. For example:
\begin{enumerate}[label=(\alph*)]
\item Mozilla Persona: The only requirements to become an IdP are to (1) own a domain name; (2) place a text file on the web server (as explained in Section~\ref{sec:ov_persona}) containing metadata such as the URL for the user authentication endpoint; and (3) assign user identifiers formatted as a standard e-mail address  \textsf{userid@domain.com} (it is not required for the IdP to actually run an e-mail server). Any website can become a Persona SP by implementing the protocol as described in Section~\ref{sec:ov_persona} (no registration with a central authority is required).
\item OpenID 2.0: Similar to Persona, becoming an IdP requires only the implementation of the protocol as described in Section~\ref{sec:ov_oid2}. No registration with any central authority is required, neither for IdPs nor for SPs. In OpenID 2.0, user identities are in the format of a URL (e.g., a profile page) over which the user proves their control by authenticating through the IdP that hosts the URL.
\end{enumerate}

A1 protocols provide \emph{\noidpvetting} and \emph{\nospsponsor}: Together, these benefits provide the deployability benefit that all SPs and IdPs can implement the protocol without requiring co-ordination. Users also have the freedom to either create an account on an existing IdP or to establish their own IdPs if they wish to do so.

However, A1 protocols lack \emph{\signalsloa: } In an implicit SP-IdP association model, SPs cannot determine users' means of authentication---even if IdPs were to provide the information, SPs cannot rely on it without a pre-established trust relationship.

\textbf{A2: Explicit association (IdP-managed namespace).} This is the most common form of web SSO today. SPs must establish an explicit association with each IdP that they wish to support, typically through a manual registration process that involves key exchange or the establishment of a shared secret. OAuth 2.0 and OpenID Connect work in this manner.

A2 protocols provide \emph{\noidpvetting}: Any entity can implement the protocol and become an IdP, thereby facilitating deployment.

However, A2 protocols lack \emph{\nospsponsor}: Each SP must register with each IdP with which it would like to associate. This typically involves a manual procedure through which the SP submits a form to establish certain parameters with the IdP, such as an application ID and a shared secret. Typically, a human administrator from the SP must manually submit the form, and the confirmation/approval by the IdP may or may not be automated. This manual process is a deployability drawback that results in SPs supporting a smaller set of IdPs, thereby raising the barrier to entry for potential users (since they may be required to obtain a new IdP account, if their own IdP is not supported).

\textbf{A3: Federated association (IdP-managed namespace).} This refers to schemes whereby a central authority (e.g., in OpenID Connect terminology, the \emph{federation operator}) maintains authoritative metadata as to which IdPs and SPs are part of the federation. Both IdPs and SPs must undergo a registration process to join the federation, but not explicitly with each other. Upon joining, SPs may rely on the authentication service of any IdPs that are part of the federation, without needing to manually pre-register with any of them. Member IdPs and SPs may be required to follow certain privacy- and security-related policies, e.g., with regards to storage of user data, or the method of authentication used. Different federations may be established between organizations based on some shared attributes (e.g., geographical area) or objectives (e.g., educational institutions). For example, CANARIE~\cite{canarie} operates a federation of research and educational institutions across Canada.

Currently, SAML (a popular implementation of which is Shibboleth) is the most popular protocol that falls within A3. OpenID Connect also has an optional federation component that is not implemented by any of the current major IdPs, namely Google, Microsoft, Yahoo, and Paypal---SPs must therefore manage independent associations with each IdP. Mobile Connect, administered by the GSM Association, is a recent extension of the OpenID Connect standard that implements a federated association model. Mobile phone service operators across the world that are members of the GSM Association are eligible to be IdPs within the federation. Therefore, once fully deployed, Mobile Connect SPs will be able to authenticate any user worldwide that has mobile phone service.

Aside from maintaining IdP and SP metadata, a principal role of the federation operator is to provide a \textbf{discovery service} through which users can be redirected to the appropriate IdP for authentication, e.g., SAML may ask the user to select their IdP manually when logging in, or Mobile Connect may determine the user's IdP automatically by their mobile phone number.

A3 protocols lack \emph{\noidpvetting}, since the federation must be managed by an operating entity, with which IdPs must register. They also lack \emph{\nospsponsor}, since SPs must manually register (and potentially abide by certain security policies) to become a member of the federation.

A3 protocols may offer \emph{\signalsloa}, if the federation operator imposes requirements for the type of authentication mechanisms used. For example, Mobile Connect grades the user's LoA on a 4-point scale, based on the authentication mechanism used.

\textbf{A4: Bridged association (federation-operator-managed namespace).} These are similar to A3 schemes in that SPs associate with IdPs via a federation operator. However, the federation operator acts as both a discovery service and an \emph{IdP bridge}, in that the SP relies on it to verify users' control over accounts under the namespace of a multitude of IdPs (as opposed to the SP communicating directly with IdPs). An example is the Mozilla Persona fallback protocol (the main Mozilla Persona protocol falls under A1 as previously discussed), in which a fallback IdP performs this service in two ways: (1) via OAuth 2.0, for several major e-mail service providers that support it; or (2) by sending a verification code to the user's e-mail address to verify that they can receive the code. In other A4 schemes, such as SAW, the SP may implement its own IdP bridge---e.g., each SAW SP verifies user control of an e-mail address by e-mailing a verification code upon every authentication attempt. A4 schemes may be tailored towards a number of different goals, namely, 
\begin{itemize}
  \item Leveraging an existing protocol deployment (e.g., SMTP and OAuth 2.0, in the case of the Persona fallback protocol) to build a new identity protocol over it.
  \item Enhancing privacy by implementing the IdP bridge as an anonymizing proxy. For example, SecureKey Concierge is an A4 protocol that provides a ``triple-blind'' service, wherein SPs are blind to users' IdP information, IdPs are blind to users' SP information, and SecureKey itself only processes unique but anonymous identifiers for users \cite{sktrustfw}.
  \item Allowing users to switch between IdPs within the federation without imposing any burden on the SP. SecureKey Concierge provides this feature, by allowing their users to switch between different banking institutions.
\end{itemize}

Third-party namespace verification may be performed in a way such that it can be validated by multiple parties (e.g., both the IdP and SP). For example, Keybase~\cite{keybase} is a directory service that maps users' public keys to their social media identities (while Keybase is not an SSO service, its function relates to third-party namespace verification). Users post their public keys on their Keybase profile page, and post corresponding cryptographically signed statements on their online accounts (e.g., Twitter, Facebook, Github) to link their online identities. Since the cryptographically signed statements are publicly-accessible, they can be verified not only by Keybase but by anybody---this reduces the trust that must be placed in Keybase itself to verify the statements. In contrast, the Mozilla Persona fallback IdP ties user identity to control of an e-mail address, which involves sending the user an e-mail that only the user can access---a process that is not publicly verifiable.

\textbf{A5: Non-attested credential manager (SP-managed namespace).} These schemes provide repositories in which users save their SP-specific credentials, which may consist of e.g., user names along with passwords or cryptographic keys. The users' credentials serve as pseudonymous identifiers, and are not associated with control over any IdP-controlled namespace. Password managers are the most common type of A5 protocols currently in use. We call A5 schemes non-attested, since SPs cannot cryptographically verify any information about the authentication mechanism between the user and IdP, including which IdP (if any) was used.

A5 schemes cannot provide \emph{\authidps} or \emph{\signalsloa}, since SPs have no means of determining the user's IdP. They provide \emph{\noidpvetting} and \emph{\nospsponsor}, since users can use any A5 scheme without obtaining SP approval. 

\textbf{A6: Attested credential manager (SP-managed namespace).} These schemes are similar to A5, but provide additional security guarantees. Through a federation authority, SPs may verify information such as the user's IdP and the type of user-to-IdP authentication used. For example, FIDO UAF uses hardware authenticator devices that act as users' IdPs by storing a unique cryptographic key pair for each SP website. Certified hardware authenticators possess an attestation key certified by the FIDO Alliance, thereby providing assurance to SPs that the authenticator can securely store private keys in hardware (precluding theft by malware) and signalling the LoA based on the type of user-to-IdP authentication that is used (e.g., PIN or biometric).

A6 protocols can provide \emph{\signalsloa}, since IdPs undergo a certification process by the federation operator, which can then signal the LoA information to SPs.

To support increased device compatibility, FIDO UAF also supports software authenticators lacking the ability to protect cryptographic keys from malware, and therefore performing no attestation (however, this is explicitly signalled). FIDO UAF is therefore a hybrid system supporting both ``strong'' attested hardware authenticators certified by a central authority (i.e., A6 scheme that provides LoA signalling) and ``weak'' unattested software authenticators (i.e., A5 protocol). We analyze hardware and software authenticators separately, since they differ substantially in the benefits offered.

\subsection{\assertauth} \label{sec:assertauth}
The following categories classify SSO schemes based on how IdPs convey user identity to SPs.

\textbf{G1: IdP assertion.} The IdP generates an identity assertion each time the user needs to access their SP. OAuth, OpenID Connect, Mobile Connect, and Shibboleth fall within this category. The IdP typically communicates the assertion to the SP through the browser via HTTP redirection, HTTP POST requests, and/or cross-origin message-passing between iframes with postMessage \cite{postmessage}.

\textbf{G2: Browser assertion.} The IdP delegates the authority for identity assertion generation to the user's browser. Mozilla Persona falls within this category: The browser generates a cryptographic key pair to sign its own self-generated identity assertions. Identity assertions also contain an IdP signature over the browser's public key, thereby allowing SPs to validate assertions. The assertion is typically communicated from client-side code to the SP. Mozilla Persona was intended to eventually be built into the browser, but as an interim measure it used a client-side JavaScript library.

\textbf{G3: User-to-SP authentication.} In this category, there is no vouching mechanism that can signal to the SP that the user has already authenticated itself to the IdP, and therefore the user must authenticate directly to the SP. The IdP may facilitate or may automate the process by which users authenticate to SPs (e.g., by storing and synchronizing the user's credentials across multiple devices, and automatically filling in forms to enter the credentials). All A5 protocols fall within this category. A4 protocols may also fall within this category, if the SP acts as its own IdP bridge (e.g., SAW).

\textbf{G4: IdP-proxied authentication.} In this case, the IdP serves as an HTTP proxy between the user and SP, which may convey identity information and/or credentials over a direct network connection to the SP. Thus, all traffic between users and SPs (both during authentication and any subsequent traffic during the authenticated session) flows through the IdP. Web proxies such as \emph{EZproxy}~\cite{ezproxy} are commonly used by educational institutions to allow off-campus access of academic material---e.g., \textsf{ieeexplore.ieee.org} may be accessed through \textsf{ieeexplore.ieee.org.edu-proxy.com}, where \textsf{edu-proxy.com} is the URL of the institution's proxy. However, such proxies typically do not convey any identity information, since the SPs operate on IP address based access control---thus, any user visiting from within the institution's IP address range is granted access. We are not aware of any proxy-based SSO schemes used in practice, but they have been proposed in literature; Impostor~\cite{pashalidis_impostor_2004} falls within this category.

G1 schemes are not \emph{\restempoutage}, since SPs must obtain an IdP-generated identity assertion to establish an authenticated session. G2 schemes \emph{may} partially provide this benefit, if the client possesses a relatively long-lived assertion-signing key; with Mozilla Persona, the expiration time of the client's signing key is set by the IdP and could be set to, e.g., 24 hours (an excessive expiration time poses additional impersonation risks if the signing key is stolen). G3 schemes \emph{may} fully provide this benefit if the credentials are cached client-side---users may thus continue to use their locally-cached credentials, but would not be able to update/synchronize any of their devices with new or updated credentials for as long as the IdP remains unreachable. G4 schemes are not \emph{\restempoutage}, since all communication between clients and SPs must be proxied through an IdP.

\subsection{\assertchan}
The methods of verifying an identity assertion conveyed from an IdP to an SP can be categorized as follows:

\textbf{C1: IdP Query.} The SP must query the IdP to determine the validity of an authentication token presented by the user (e.g., to verify that the token was issued by the IdP, intended for use by the SP in question, and has not expired).

\textbf{C2: Local Verification.} The SP itself directly verifies that the user has correctly authenticated. This may be done if (1) the user authenticated directly to the SP (see G3), or (2) if the identity assertion is cryptographically signed by the IdP and can be verified by the SP.

The OpenID 2.0 specification supports both C1 and C2, but recommends C2 since it reduces the number of required protocol round trips \cite{openid2}---C1 is called \emph{direct validation}, and C2 is called \emph{association mode}. Association mode requires a Diffie-Hellman key exchange to establish a shared key between the IdP and SP; the IdP uses the key to compute a MAC on identity assertions destined for the associated SP. Association mode prevents tampering of identity assertions via MITM attacks that may occur if direct validation is used. However, TLS also prevents tampering, and is recommended (but not required) by the specification. 

Mozilla Persona supports only C2: IdPs are \emph{stateless}, since they do not generate any identity assertions, and are not informed upon the generation of any identity assertions. Instead, the browser generates the identity assertion (see \ref{sec:assertauth}), which contains (1) the SP domain for which the assertion was generated, (2) the assertion's expiry time, (3) a client-generated cryptographic signature of the assertion, and (4) an IdP-generated signature of the client-generated signing key. The SP can validate assertions locally, since the only information required from the IdP is the IdP's public key (see Section~\ref{sec:ov_persona}), to be used for verifying the IdP-generated signature of the client's signing key. All other required information for validating the assertion (e.g., SP domain, assertion expiry time) is present within the assertion itself.

SAML and the OAuth 2.0 family of protocols (which includes OpenID Connect and Mobile Connect) are \emph{stateful} protocols: Clients must either (1) obtain a bearer token from the IdP and send it to the SP, which must then query the IdP to determine the assertion's validity (known as \emph{Implicit Grant} in OAuth 2.0); or (2) obtain a temporary code from the IdP and send it to the SP, which must then exchange it for a token from the IdP (known as \emph{Authorization Code Grant} in OAuth 2.0).

C2 protocols that are stateless can provide \emph{\privatefromidp} (this includes all A5 and A6 protocols), since IdPs are not aware of the SPs with which users are associated. Protocols that are both C1 and A4 may also offer this benefit if blinding is used (see Section~\ref{sec:idpspassoc}).

\subsection{\machinetype} \label{sec:machinetype}
\begin{figure}
\centering
\includegraphics[width=\columnwidth]{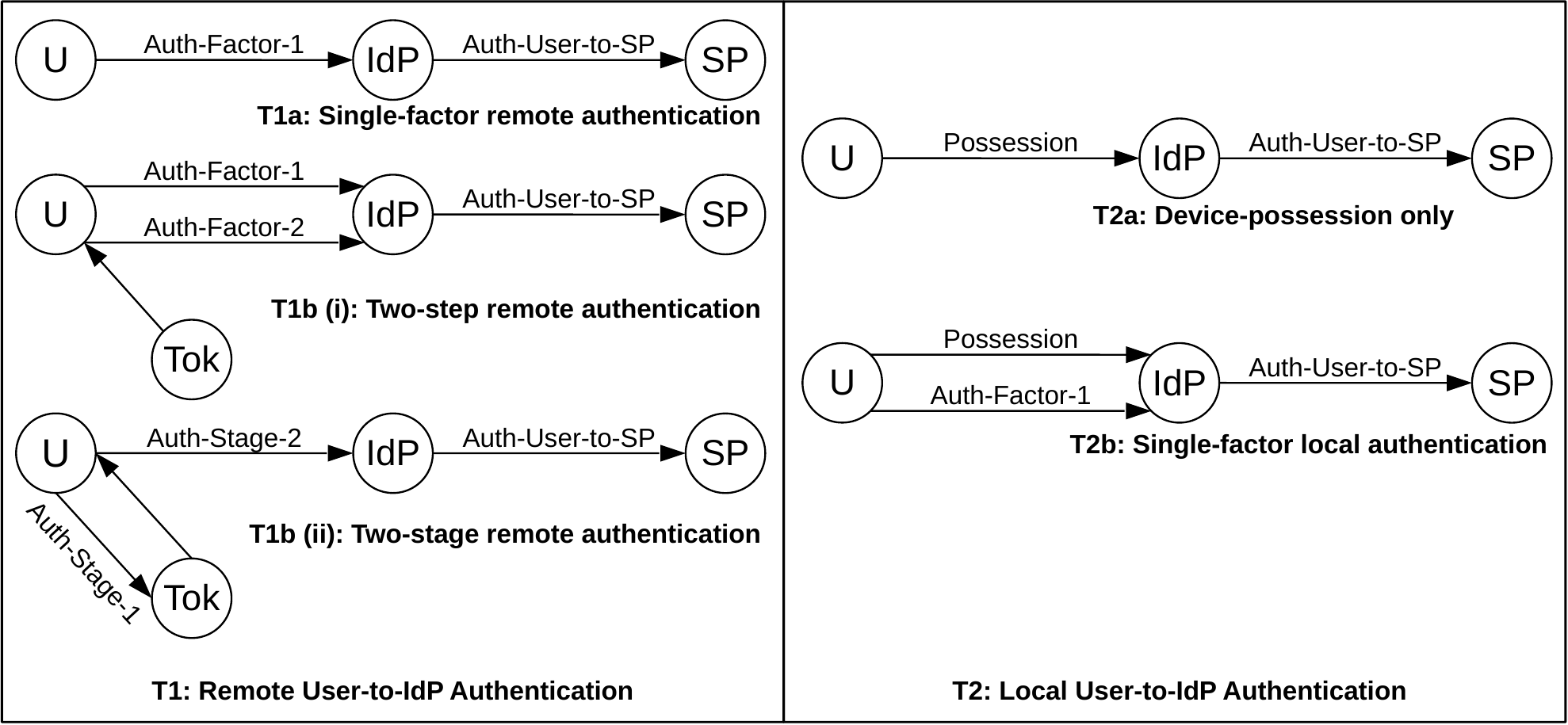}
\caption{Authentication flow for single-factor and two-factor T1 (Remote authentication) and T2 (Local authentication) schemes. \textbf{U}, \textbf{IdP}, \textbf{SP}, and \textbf{Tok} represent the user, identity provider (remote or local), service provider, and hardware token, respectively.\label{fig:twostage}}
\end{figure}

Single sign-on involves both a user-to-IdP component and an IdP-to-SP component---the latter was covered in Section~\ref{sec:assertauth}. The user-to-IdP component can be classified into two general categories based on whether the IdP is local (i.e., a device under physical control of the user) or remote (i.e., a remote server). Evaluating specific user-to-web authentication schemes \cite{bonneau2012quest} is out of the scope of this paper. However, we highlight the differences between different categories of single-factor and two-factor schemes, particularly for FIDO UAF and Mobile Connect. The following two categories are illustrated in Fig.~\ref{fig:twostage}.

\textbf{T1: Remote authentication.} The user authenticates to a remote IdP, by means of a password or any other form of user-to-web authentication. This may involve one or more authentication factors:
\begin{enumerate}[label=(\alph*)]
\item \textbf{Single factor.} Most currently-deployed IdPs that use single-factor authentication use conventional password-based authentication. One exception is Mobile Connect with LoA2, which validates the user's possession of their mobile phone. Depending on the implementation, this may require the user to type in an SMS OTP received on their phone, or the Mobile Connect authenticator application may display a popup on the phone for the user to confirm the authentication request by tapping an ``OK'' button. \item \textbf{Two factor.} Two-factor authentication can be further divided into two sub-categories, that we define as follows (to avoid overloading or redefining the term ``factor''):
\begin{enumerate}[label=(\roman*)]
\item \textbf{Two-step.} Two-factor authentication in its most popular form on the web involves two steps, wherein the server verifies the user's password (\emph{what-you-know}) in addition to their possession of a hardware token (\emph{what-you-have}) such as a smartphone or USB dongle. Here, the two factors are independent from each other.
\item \textbf{Two-stage.} Mobile Connect with LoA3 requires that the mobile device (\emph{what-you-have}) be protected by a local authentication mechanism such as a PIN or biometric. We call this two-stage, since the first authentication stage (e.g., entering a PIN) occurs on the device itself, with no participation from the IdP. Only the second stage (e.g., receiving an SMS OTP or tapping an ``OK'' button as described above) involves the IdP. Here, the second factor (stage) depends on the first.
\end{enumerate}
\end{enumerate}
The security offered by the first stage of two-stage authentication depends on the implementation of the local authentication stage. For example, a local authentication mechanism that relies on a secure boot mechanism and trusted execution environment to render a stolen device inoperable after ten consecutive failed authentication attempts would offer higher security than a device on which an attacker is given unlimited authentication attempts or can easily extract the plaintext secret from memory.

\textbf{T2: Local authentication.} The user authenticates to an IdP hosted on a local device under physical control of the user. This differs significantly from T1 schemes that use hardware authenticators (e.g., Mobile Connect or OpenID Connect with Google two-step verification) to authenticate to a remote IdP. Local authentication can be further classified as follows:
\begin{enumerate}[label=(\alph*)]
\item \textbf{Device-possession only.} No user-to-device authentication is required, aside from device possession (typically implied for local authentication). Firefox Sync 1.5 and 2.0 are both in this category by default, since the password vault is stored on-disk in plaintext and can easily be extracted when in possession of the device. The two protocols differ in their cross-device synchronization mechanisms, which is distinct from the local user-to-IdP authentication mechanism discussed here.
\item  \textbf{Single factor.} A password or other user-to-device authentication scheme (e.g., smart card, PIN, biometrics) is used to authenticate to the on-device local IdP. FIDO UAF is in this category since it requires, e.g., a PIN or biometric to be used locally before the user can be authenticated to any remote SPs. Firefox Sync could be classified into this category if, e.g., the password vault is protected by a master password or by full-disk encryption. However, weaker mechanisms such as the user login mechanism in many commodity desktop operating systems that do not encrypt user data can be easily defeated---for example, an attacker in physical possession of the device could circumvent the user login (e.g., by removing the hard disk and reading its contents) to extract the unencrypted password vault from the device's storage. Therefore, similarly to the first stage of two-stage remote authentication as discussed above, the security offered by a single-factor local authentication scheme depends on implementation and device configuration details, e.g., use of secure boot and full-disk encryption.
\end{enumerate}
Local authentication may use two or more factors in addition to the possession factor, but this is not typically used (we are not aware of any such scheme used in practice).

T1 offers \emph{\nodevicepair}, since users can authenticate to a remote IdP from any device; T2 cannot offer this benefit, since users must configure their new devices before they are able to authenticate to SPs through them.

\begin{table*}[ht]
\renewcommand{\arraystretch}{1.0}
\small
\centering
\scalebox{0.85}{
\begin{tabular}{|L{0.24\linewidth}|L{0.3\linewidth}|L{0.5\linewidth}|}
\hline
\textbf{Authentication Type} & \textbf{Possible Threat} & \textbf{Corresponding Mitigation}
\\\hline

\textbf{T1a.} Single factor, remote IDP &
Attacker may break first factor &
Add second factor
\\\hline

\multirow{3}{1.5in}{\textbf{T1b(i).} Two-step, remote IdP} &
\multirow{3}{1.75in}{Attacker may breach remote IdP and steal server-side\\
stored credentials, e.g., password hash (to perform offline guessing attack) and OTP seed} &
Use non-replayable challenge-response authentication (e.g., public-key crypto) \\\cline{3-3}
& & Use hardware security mechanisms to securely store replayable server-side credentials (e.g., use hardware security module to store symmetric key and use it to perform HMAC of passwords instead of a standard hash) \\
& & \\\hline

\textbf{T1b(ii).} Two-stage, remote IdP &
Attacker in possession of hardware token may break first factor, since it is validated locally &
Use hardware security on authentication token (e.g., trusted boot, trusted execution environment, encrypted storage) to protect authentication from local attacks
\\\hline
\textbf{T2a.} Device-possession only, local IdP &
Device theft &
Add a local authentication factor
\\\hline
\multirow{3}{1.5in}{\textbf{T2b.} Single factor, local IdP} &
Attacker may circumvent local authentication &
Use hardware security on authentication token (e.g., trusted boot, trusted execution environment, encrypted storage) to protect authentication from local attacks
\\\cline{2-3}
& \multirow{2}{1.75in}{Attacker may breach local\\ IdP (e.g., via malware) and steal stored credentials to impersonate user} &
Use non-replayable challenge-response authentication (e.g., public-key crypto) and store credentials in secure hardware enclave \\\cline{3-3}
& & Ensure that replayable credentials are never exposed to any party other than the corresponding SP, e.g., by using mandatory access control to ensure that only the IdP has access to stored credential and only releases them to authorized SPs
\\\hline
\end{tabular}}
\caption{Threats and corresponding mitigations to discussed user-to-IdP authentication types. Single-factor user-to-device authentication (\textbf{T2b}) requires user possession of the device, and therefore provides security benefits similar to two-factor user-to-remote-IdP authentication (\textbf{T1b}). \label{table:authfactors}}
\end{table*}

\subsubsection{Trusted Computing}

Secure hardware key storage, and trusted execution environments \cite{murdoch_tee} can be used to enhance the security of both T1 and T2 schemes. The primary benefit is the reduction of damage that can be done by an attacker that breaches the IdP (e.g., via a remote exploit, malware, or physical device theft)---both for IdPs hosted on a remote server or on a local device under the user's control.  For context, Table~\ref{table:authfactors} summarizes some attacks against user-to-IdP authentication along with corresponding defenses.

Another function of trusted computing in SSO is that it allows one party to make cryptographically-verifiable guarantees to another party. For example, through trusted execution, FIDO UAF attested hardware authenticators guarantee to SPs that the user has been authenticated on the local device via, e.g., a physical biometric. Mobile Connect can use the mobile phone's SIM card as a secure element for credential storage and as a trusted execution environment to authenticate the user, guaranteeing a higher LoA \cite{gsma_cpas04}. SSO may also benefit directly or indirectly from other applications of trusted computing that enhance the security and privacy characteristics of online services. For example, a remote server with Intel SGX can guarantee \cite{sgxpasswords_2017} that the password entered by the user will be processed in a secure enclave and that only the CMAC (cipher-based MAC) will be stored server-side, thereby eliminating the possibility of offline attacks in the event that the password database is leaked. The secure messaging application Signal \cite{signal_private_discovery} uses SGX for private contact discovery, allowing clients to poll a secure enclave on the server to discover which of their contacts are also using Signal, without revealing the user's contact list to the server.

\subsection{\devauthtoken} \label{sec:devauthtoken}
The means by which users can access their SP accounts from multiple devices is related to, but not fully determined by, the user-to-IdP authentication type (discussed above in Section~\ref{sec:machinetype}). Schemes can be categorized by this property as follows.

\textbf{M1: Device-independent.} This includes remote authentication schemes (T1) that do not rely on a hardware authenticator device that the user needs to carry.

\textbf{M2: Portable hardware token.} This includes remote authentication schemes (T1) that use a portable hardware authenticator device, e.g., MobileConnect (both single-factor LoA2 and two-factor LoA3).

\textbf{M3: Per-device authorization.} This includes schemes in which each of the user's devices need to be individually authorized by the user's IdP (for T1 schemes) or the user's SPs (for T2 schemes). Mobile Connect (T1) could be extended to allow mobile network operators to authorize multiple SIM cards that users could install in their devices (e.g., laptops); this would result in a hybrid of M2 and M3, since users could authorize their devices equipped with a cellular modem, but would need to carry around their mobile phone to log in on non-equipped devices. FIDO UAF (T2) is an M3 scheme, since users that obtain a new FIDO-enabled device need to authorize the new device with each of their SPs.

\textbf{M4: Device pairing.} This includes local authentication schemes in which new user devices need to be authorized by (i.e., ``paired'' with) an existing user device. This could be achieved by, e.g., securely transferring a symmetric key from one device to another, in the case of Firefox Sync. A more complex mechanism may be to generate a new cryptographic key pair on the new user device, and authorizing the new key pair from the existing device (e.g., by signing the public key).

M1 cannot offer \emph{\clientleak}, but offers \emph{\nohwtoken}. M2 may offer \emph{\clientleak} but cannot offer \emph{\nohwtoken}. M3 may offer \emph{\clientleak} but M4 can only offer it if public-key cryptography is used and the private key cannot be extracted from the device. Neither M3 nor M4 can offer \emph{\nodevicepair}, since users cannot authenticate from a new device to their SP without first enrolling the device onto their SP accounts.

\begin{table*}[htp]
\small
\centering
\setlength{\tabcolsep}{2pt}
\scalebox{0.93}{
\begin{tabular}{l rllll@{\hskip 0.25in} rlll@{\hskip 0.25in} rll@{\hskip 0.25in} rlll@{\hskip 0.25in} rll}

\headrowx{}{90} & \headrowx{\idpspassoc}{90} & \headrowx{\assertauth}{90} & \headrowx{\assertchan}{90} & \headrowx{\machinetype}{90} & \headrowx{\devauthtoken}{90} &
\headrowx{B1: \portableid}{90} & \headrowx{B2: \nodevicepair}{90} & \headrowx{B3: \nohwtoken}{90} & \headrowx{B4: \restempoutage}{90} & \headrowx{B5: \noidpvetting}{90} & \headrowx{B6: \nospsponsor}{90} & \headrowx{B7: \spusersecret}{90} & \headrowx{B8: \deviceboundsecret}{90} & \headrowx{B9: \signalsloa}{90} & \headrowx{B10: \authidps}{90} & \headrowx{B11: \notpimpersonation}{90} & \headrowx{B12: \privatefromidp}{90} & \headrowx{B13: \spunlinkable}{90} & \headrowx{B14: \nodatasharing}{90} 
\\ \hline

Scheme &
\multicolumn{5}{c}{\hspace{-0.125in}Design Properties} &
\multicolumn{4}{c}{\hspace{-0.125in}Usability} &
\multicolumn{3}{c}{\hspace{-0.125in}Deployability} &
\multicolumn{4}{c}{\hspace{-0.125in}Security} &
\multicolumn{3}{c}{\hspace{-0.125in}Privacy} 
\\ \hline

OpenID 2.0
& A1 & G1 & C2 & T1 & M1 &
  & \full & \full &  & \full & \full & \full & \footnotesize{b} &  & \full &  &  &  & \full 
\\

Mozilla Persona	&
A1 & G2 & C2 & T1 & M1 &
  & \full & \full & \prt & \full & \full & \full & \footnotesize{b}  &  & \full &  & \full &  & \full 
\\

OAuth 2.0 &
A2 & G1 & C1 & T1 & M1 &
  & \full & \full &  & \full &  & \full & \footnotesize{b} &  & \full &  &  &  &  
\\
  
OpenID Connect &
A2 & G1 & C1 & T1 & M1 &
  & \full & \full &  & \full &  & \full & \footnotesize{b} &  & \full &  &  &  &  
\\ 

Mobile Connect &
A3 & G1 & C1 & T1 & M2 &
  & \full &  &  &  & \full & \full & \full & \full & \full &  &  & \full &  
\\

Shibboleth &
A3 & G1 & C1 & T1 & M1 &
  & \full & \full &  &  & $~\S$ & \full & \footnotesize{b} & \full & \full &  &  & $~\S$ &  
\\   

SAW &
A4 & G3 & C1 & T1 & M1 &
\prt & \full & \full &  & \full & \full & \full & \footnotesize{b} &  & \full &  &  &  & \full 
\\
						
Mozilla Persona Fallback &
A4 & G2 & C2 & T1 & M1 &
  & \full & \full & \prt & \full & \full & \full & \footnotesize{b} &  & \full & $~\dagger$ & \full & & \full 
\\

SecureKey Concierge &
A4 & G1 & C1 & T1 & M1 &
\prt & \full & \full &  &  & \full & \full & \footnotesize{b} & \full & \full & $~\dagger$ & \full & \full & \full 
\\\hline

Firefox Sync 1.5 &
 A5 & G3 & C2 & T2a & M4 &
\full &  & \full & \full & \full & \full &  & \footnotesize{c} &  &  & \full& \full & \full & \full 
\\

Firefox Sync 2.0 &
A5 & G3 & C2 & T2a & M4 &
\full & \prt & \full & \full & \full & \full &  &  &  &  & \prt& \full & \full & \full 
\\

Impostor &
A5 & G4 & C2 & T1 & M1 &
  & \full &  &  & \full & \full &  &  &  &  &  &  & \full & \full 
\\

FIDO UAF (Non-Attested) &
A5 & G1 & C2 & T2b & M3 &
\full &  & \full & \full & \full & \full & \full & \footnotesize{bc} &  &  & \full& \full & \full & \full 
\\

FIDO UAF (Attested) &
A6 & G1 & C2 & T2b & M3 &
  &  & \full  & \full &  & \full & \full & \full & \full & \full & \full& \full & \full & \full 
\\

\hline
\end{tabular}
}
\caption{SSO schemes (rows) categorized across design properties (columns, left half) and evaluated across benefits (columns, right half) from Section~\ref{sec:sso_properties} (sorted by IdP-SP Association Model). FIS schemes are listed above the dividing line and CM schemes are in the lower portion; recall that FIS schemes are A1-A4, CM schemes are A5-A6. Bullets represent benefits provided; empty circles represent partially-provided benefits; empty cells represent benefits not provided; letters in column B8 represent sub-benefits provided by schemes that do not offer all three sub-benefits. ${}^\dagger$Susceptible to impersonation by either of \emph{two} third parties. ${}^\S$Offered via optional protocol feature. \label{table:sso_categorization}}
\end{table*}
 
\section{Evaluation of SSO Schemes} \label{sec:sso_compare}
Here, we explain our evaluation (based on the benefits discussed in Section~\ref{sec:benefits}) and categorization (based on the design properties discussed from Sections~\ref{sec:idpspassoc} through \ref{sec:machinetype}) of each SSO scheme introduced in Section~\ref{sec:sso_protocols}, as is summarized in Table~\ref{table:sso_categorization}.

\textbf{Firefox Sync.}
We evaluate Firefox Sync 1.5 and 2.0 together, and indicate differences between the two versions in the benefits offered. Firefox Sync offers \emph{\portableid}, since the user's device is the IdP, and transferring credentials to new devices is supported via a synchronization mechanism. Users can authenticate to SPs from any synchronized device without an additional hardware token (\emph{\nohwtoken}), even when the synchronization server is down (\emph{\restempoutage} for authenticating to SPs, but not for updating or adding new passwords to the vault). Firefox Sync 1.5 does not offer \emph{\nodevicepair}, since users must ``pair'' any new devices with one of their existing devices by transferring their symmetric key for decrypting their password vault; Firefox Sync 2.0 partially offers this benefit, since the decryption key is derived from the user's password and must be entered during the device setup process.

Users' devices are not vetted by any central authority (\emph{\noidpvetting}) and SPs do not require any IdP-specific server-side modifications (\emph{\nospsponsor}). However, SPs must still store and protect per-user secret credentials (\emph{\spusersecret} not offered). \emph{\signalsloa} and \emph{\authidps} are not provided since SPs cannot distinguish between IdPs.

Firefox Sync 1.5 offers \emph{\notpimpersonation} since the password vault is locally encrypted on the user's device with a randomly-generated symmetric key; Firefox Sync 2.0 partially offers this benefit, since the vault encryption key is password-derived and is therefore subject to offline attack by the synchronization server. Firefox Sync 1.5 does not offer \emph{\clientleak}, since the vault encryption key is not stored in hardware and can be transferred to all the user's devices via the pairing process; neither does Firefox Sync 2.0, since it derives the key from a password, which may be captured by an attacker. Neither offer \emph{\spleak}, since SPs must store user passwords server-side (hashed passwords may still be cracked via offline guessing attacks). Firefox Sync 1.5 offers \emph{\remoteleak}, since password vaults stored on the synchronization server are encrypted with randomly-generated keys known only to clients, and therefore are infeasible to decrypt if leaked; Firefox Sync 2.0 partially offers this benefit, since it relies on password-derived keys, and therefore leaked password vaults may be subject to offline attack (with attack success depending on the strength of the user-chosen passwords).\footnote{This grading depends on the assumption that passwords are not re-used across SPs. Password re-use results in the loss of the \emph{\remoteleak} benefit, since a leak from one SP can lead to account compromise on other SPs.}

Firefox Sync offers \emph{\privatefromidp}, \emph{\spunlinkable}, and \emph{\nodatasharing} since all authentication is performed through the user's device without any participation by a remote third-party server.

\textbf{FIDO UAF.}
We evaluate the attested and non-attested FIDO UAF variants together, and indicate differences between the two versions in the benefits offered. UAF offers \emph{\nohwtoken} and \emph{\restempoutage}: users can access their SP accounts from any of their UAF-enabled personal devices without dependence on any third-party remote servers, but users must first set up UAF on their devices with each individual SP (so \emph{\nodevicepair} is not offered). Unattested UAF offers \emph{\portableid} since users can transfer their credentials to a new device, but attested UAF does not offer this benefit since keys are stored securely in the device's hardware and cannot be extracted.

Non-attested UAF offers \emph{\noidpvetting}, but attested UAF does not, since IdPs must have their attestation key signed by the FIDO Alliance. Both UAF variants offer \emph{\nospsponsor} since SPs do not require any IdP-specific server-side modifications---SPs need only possess the FIDO Alliance root certificate to verify the validity of any FIDO-signed IdP attestation keys. Both variants offer \emph{\spusersecret}, since SPs store users' public keys.

Attested UAF offers \emph{\clientleak}, since private keys are stored in hardware and cannot be extracted; non-attested UAF does not, since keys are stored in software and are thus susceptible to capture. Both variants offer \emph{\spleak} and \emph{\remoteleak}, since verifying servers only store users' public keys (which are not secret). Attested UAF allows SPs to validate signed IdP attestation certificates, which indicate the method of authentication used (e.g., 4-digit PIN or biometric) and the hardware vendor and model number (allowing SPs to phase out support for hardware authenticators in which vulnerabilities have been found), thereby offering both \emph{\signalsloa} and \emph{\authidps}; non-attested UAF does not offer these benefits, since they are not certified by FIDO and therefore use self-signed attestation certificates, with the signing key stored unprotected in software. Both UAF variants offer \emph{\notpimpersonation}, since users' authentication keys are only stored on their own devices.

Both UAF variants offer \emph{\privatefromidp}, \emph{\spunlinkable}, and \emph{\nodatasharing} since all authentication is performed through the user's device without any participation by a remote third-party server. To limit linkability between SP accounts, attested UAF must use an attestation certificate that is shared by at least 100,000 hardware devices (the same batch/revision of any hardware authenticator model shares the same attestation certificate).

\textbf{Impostor.}
\emph{\portableid} is not offered, since users cannot change IdPs without updating all of their SP accounts with their new identity. \emph{\nodevicepair} is offered but \emph{\nohwtoken} is not offered (assuming a hardware OTP token is used for user-to-IdP authentication; see Section~\ref{sec:ov_shibboleth}). \emph{\restempoutage} is not offered, since users cannot authenticate to SPs without first authenticating through their IdP.

\emph{\noidpvetting} is offered, since any domain owner can set up their own IdP. \emph{\nospsponsor} is offered, since users can provide any e-mail address as their user name to SPs. \emph{\spusersecret} is not offered, since SPs store user passwords. 

\emph{\clientleak} is offered, since Impostor uses a challenge-response authentication protocol designed for use on untrusted access devices, without exposing any long-term credentials to them. \emph{\spleak} is not offered, since SPs must store user passwords server-side (hashed passwords may still be cracked via offline guessing attacks). \emph{\remoteleak} is not offered, since passwords are stored in plaintext by IdPs and are susceptible to capture and reuse. \emph{\signalsloa} and \emph{\authidps} are not offered, since there is no SP-IdP communication aside from automated password submission. \emph{\notpimpersonation} is not offered, since IdPs are in possession of users' SP passwords (unless users self-host an IdP).

\emph{\privatefromidp} is not offered, since users access all their SPs through the IdP proxy. \emph{\spunlinkable} and \emph{\nodatasharing} are offered, since IdPs only communicate SP-specific user names and passwords.

\textbf{SAW.} 
SAW partially offers \emph{\portableid}, since users can use e-mail forwarding on an interim basis when transitioning from one address to another (but all SPs must still be individually updated). \emph{\nodevicepair} and \emph{\nohwtoken} are offered, assuming password-based authentication to the e-mail provider. \emph{\restempoutage} is not offered, since e-mail service outage prevents users from receiving OTPs.

\emph{\noidpvetting} and \emph{\nospsponsor} are offered since any domain owner can host a mail server and exchange e-mails with any other mail servers. \emph{\spusersecret} is offered since SPs do not need to store any user secrets (only their e-mail addresses).

\emph{\clientleak} is not offered (assuming password authentication); \emph{\spleak} is offered, since SPs do not store any user authentication secret; and \emph{\remoteleak} is not offered since hashed passwords stored by IdPs are susceptible to exposure and offline attack, resulting in attacker access to users' SP accounts. \emph{\signalsloa} is not offered, since the user-to-IdP authentication method is not signalled to SPs. \emph{\authidps} is offered, since SPs can whitelist or blacklist e-mail providers by domain name. \emph{\notpimpersonation} is not offered: a malicious mail server administrator can impersonate their own e-mail users.

\emph{\privatefromidp} is not offered since IdPs can track the SPs that their users visit. \emph{\spunlinkable} is not offered, assuming users use the same e-mail address across different SP accounts (users can theoretically use different e-mail aliases pointing to the same e-mail account to obtain this benefit, but we do not assume this to be the case due to its impracticality). \emph{\nodatasharing} is offered since SMTP does not provide any user information.

\textbf{OAuth 2.0 and OpenID Connect.}
The following evaluation applies to both OAuth 2.0 and OpenID Connect, since their differences are small enough in their default configurations such that they do not differ in their benefits offered.\footnote{As discussed in Section~\ref{sec:ov_oidc}, OpenID Connect offers a more strictly-defined protocol for authentication (compared to OAuth 2.0), which results in less variation between implementations, thereby enhancing interoperability. It also offers optional features that can be implemented to, e.g., establish a federated association model as is done by Mobile Connect (which is therefore evaluated separately).} \emph{\portableid} is not offered, since users cannot change IdPs without updating all of their SP accounts with their new identity. \emph{\nodevicepair} and \emph{\nohwtoken} are offered, assuming conventional password authentication. \emph{\restempoutage} is not offered, since users cannot authenticate to SPs without first authenticating through their IdP.

\emph{\noidpvetting} is provided, since any domain owner can set up their own IdP. \emph{\nospsponsor} is not offered, since SPs must complete a manual registration process with each individual IdP. \emph{\spusersecret} is offered, since SPs only store a \{user ID, IdP ID\} pair.

\emph{\clientleak} is not offered (assuming password authentication); \emph{\spleak} is offered, since SPs do not store any user authentication secret; and \emph{\remoteleak} is not offered since hashed passwords stored by IdPs are susceptible to exposure and offline attack, resulting in attacker access to users' SP accounts. \emph{\signalsloa} is not offered. \emph{\authidps} is offered, since SPs must explicitly implement support for each IdP. \emph{\notpimpersonation} is not provided, since IdPs can impersonate their users.

\emph{\privatefromidp} is not offered, since user authentication requires redirecting the user's browser between the IdP and SP. \emph{\spunlinkable} is not offered, since it is not required for IdPs to assign unique pairwise (user-to-SP) pseudonymous identifiers to prevent identity correlation across SP accounts. \emph{\nodatasharing} is not offered, since OAuth 2.0 and OpenID Connect based IdPs provide SPs with extensive access to user profile information.

\textbf{OpenID 2.0.}
\emph{\portableid} is not offered, since users cannot change IdPs without updating all of their SP accounts with their new identity. \emph{\nodevicepair} and \emph{\nohwtoken} are offered, assuming conventional password authentication. \emph{\restempoutage} is not offered, since users cannot authenticate to SPs without first authenticating through their IdP.

\emph{\noidpvetting} is provided, since any domain owner can set up their own IdP. \emph{\nospsponsor} is provided, since user IDs are in the format of a globally-resolvable user profile URL via which SPs can dynamically discover the user's IdP and initiate the authentication process. \emph{\spusersecret} is offered, since SPs only store user profile URLs.

\emph{\clientleak} is not offered (assuming password authentication); \emph{\spleak} is offered, since SPs do not store any user authentication secret; and \emph{\remoteleak} is not offered since hashed passwords stored by IdPs are susceptible to exposure and offline attack, resulting in attacker access to users' SP accounts. \emph{\signalsloa} is not offered. \emph{\authidps} is offered, since SPs can filter IdPs by domain name. \emph{\notpimpersonation} is not provided, since IdPs can impersonate their users.

\emph{\privatefromidp} is not offered, since user authentication requires redirecting the user's browser between the IdP and SP. \emph{\spunlinkable} is not offered, since user IDs (i.e., profile URL) are the same across different SPs. \emph{\nodatasharing} is offered, since OpenID 2.0 does not support the exchange of user profile information.

\textbf{Mobile Connect.}
\emph{\portableid} is not offered, since users lose access to all their SP accounts upon changing their mobile network operator---the GSMA aims to address this in the future, by allowing users to transfer their identity across IdPs if they keep the same phone number when changing service providers \cite{mc_lifecycle}. \emph{\nodevicepair} is offered since users can log into their SP accounts on any device as long as they are in possession of their mobile phone (thus \emph{\nohwtoken} is not offered).  \emph{\restempoutage} is not offered, since users cannot authenticate to SPs without first authenticating through their IdP.

\emph{\noidpvetting} is not provided, since IdPs must be a GSM mobile network operator. \emph{\nospsponsor} is provided, since Mobile Connect provides a discovery service through which SPs are redirected to the correct IdP based on the user's mobile phone number. \emph{\spusersecret} is offered, since SPs only store pseudonymous user identifiers provided by IdPs.

\emph{\clientleak} is offered, since authentication private keys are securely stored in users' SIM cards; \emph{\spleak} is offered, since SPs do not store any user authentication secret; and \emph{\remoteleak} is offered since IdPs do not store any user authentication secrets (only public keys). \emph{\signalsloa} is offered, since IdPs signal SPs with an LoA for each authenticated user.  \emph{\authidps} is offered. \emph{\notpimpersonation} is not provided, since IdPs can impersonate their users.

\emph{\privatefromidp} is not offered, since user authentication requires redirecting the user's browser between the IdP and SP. \emph{\spunlinkable} is offered, since IdPs assign unique pairwise (user-to-SP) pseudonymous identifiers to prevent user identity correlation across SPs. \emph{\nodatasharing} is not offered, since Mobile Connect allows IdPs to share user attributes.

\textbf{Mozilla Persona.}
The following evaluation applies to both Mozilla Persona and Mozilla Persona Fallback, with any differences between the two variants (in terms of benefits offered) explicitly indicated. \emph{\portableid} is not offered, since users cannot change IdPs without updating all of their SP accounts with their new identity. \emph{\nodevicepair} and \emph{\nohwtoken} are offered, assuming conventional password authentication. \emph{\restempoutage} is partially offered, subject to the browser certificate expiry time set by the IdP (typically 24 hours)---the user's browser can generate signed identity assertions that can be validated by SPs (if the SP has a cached copy of the IdP public key).

\emph{\noidpvetting} is offered, since any domain owner can set up their own IdP. Persona offers \emph{\nospsponsor}, since users can provide any e-mail address as their user name to SPs. \emph{\spusersecret} is offered, since SPs only store users' e-mail addresses.

\emph{\clientleak} is not offered (assuming password authentication); \emph{\spleak} is offered, since SPs do not store any user authentication secret; and \emph{\remoteleak} is not offered since hashed passwords stored by IdPs are susceptible to exposure and offline attack, resulting in attacker access to users' SP accounts. \emph{\signalsloa} is not offered. \emph{\authidps} is offered, since SPs can filter IdPs by domain name. \emph{\notpimpersonation} is not offered, since IdPs (and also the fallback server, in the case of the Persona Fallback scheme) can impersonate their users.

\emph{\privatefromidp} is offered, since identity assertions are generated locally on users' browsers. \emph{\spunlinkable} is not offered, since user IDs (i.e., e-mail address) are the same across different SPs. \emph{\nodatasharing} is offered, since Persona does not support the exchange of user profile information.

\textbf{Shibboleth.}
\emph{\portableid} is not offered, since users cannot change IdPs without updating all of their SP accounts with their new identity. \emph{\nodevicepair} and \emph{\nohwtoken} are offered, assuming conventional password authentication. \emph{\restempoutage} is not offered, since users cannot authenticate to SPs without first authenticating through their IdP.

\emph{\noidpvetting} is offered, since any domain owner can set up their own IdP. \emph{\nospsponsor} may be offered if SPs use a federation discovery service that allows users to pick from a list of member IdPs in the federation. \emph{\spusersecret} is offered, since SPs only store a \{user ID, IdP ID\} pair.

\emph{\clientleak} is not offered (assuming password authentication); \emph{\spleak} is offered, since SPs do not store any user authentication secret; and \emph{\remoteleak} is not offered since hashed passwords stored by IdPs are susceptible to exposure and offline attack, resulting in attacker access to users' SP accounts. \emph{\signalsloa} is offered, since the underlying SAML protocol allows IdPs to signal an LoA for each authenticated user (enforcement of the feature depends on the federation authority). \emph{\authidps} is offered, since SPs can choose which IdPs to support. \emph{\notpimpersonation} is not offered, since IdPs can impersonate their users.

\emph{\privatefromidp} is not offered, since user authentication requires redirecting the user's browser between the IdP and SP. \emph{\spunlinkable} may be offered, since Shibboleth optionally supports IdP-assigned unique pairwise (user-to-SP) pseudonymous identifiers to prevent user identity correlation across SPs. \emph{\nodatasharing} is not offered, since Shibboleth supports the transfer of user profile information.

\textbf{SecureKey Concierge.}
\emph{\portableid} is partially offered, since SecureKey maintains a unique pseudonymous identifier for each user that is mapped to the user's IdP of choice, and users may switch to another IdP within the federation at any time. \emph{\nodevicepair} and \emph{\nohwtoken} are offered, assuming conventional password authentication. \emph{\restempoutage} is not offered, since users cannot authenticate to SPs without first authenticating through their IdP.

\emph{\noidpvetting} is not provided, since IdPs must be approved by SecureKey. \emph{\nospsponsor} is offered, since SecureKey can add or remove support for IdPs without requiring any server-side updates to SPs. \emph{\spusersecret} is offered, since SPs only store pseudonymous user identifiers provided by SecureKey.

\emph{\clientleak} is not offered (assuming password authentication); \emph{\spleak} is offered, since SPs do not store any user authentication secret; and \emph{\remoteleak} is not offered since hashed passwords stored by IdPs are susceptible to exposure and offline attack, resulting in attacker access to users' SP accounts. \emph{\signalsloa} is offered. \emph{\authidps} is offered, since SecureKey allows SPs to choose the set of IdPs that they wish to allow. \emph{\notpimpersonation} is not offered, since both SecureKey and users' IdPs can impersonate their users.

\emph{\privatefromidp}, \emph{\spunlinkable} and \emph{\nodatasharing} are offered: users are indirectly forwarded from SPs to IdPs via SecureKey, which only exchanges blinded identifiers (see Section~\ref{sec:ov_securekey}) to offer privacy from IdPs and prevent user identity correlation across SPs. 

\section{Discussion} \label{sec:discussion}

Among the schemes analyzed, none possesses the technical features necessary to offer all the benefits discussed. This is to be expected, since each design property discussed necessitates some trade-offs between benefits. Moreover, even for schemes that are able to offer certain benefits, conflicting interests between the stakeholders involved (namely users, SPs, and IdPs) may result in the benefits not being offered---e.g., OpenID Connect supports the use of pseudonymous identifiers, which provides \emph{\spunlinkable}, but major IdPs such as Google do not implement this feature \cite{google_oidc}. This motivates a discussion to examine the scenarios under which each individual scheme offers the most appropriate combination of benefits.

Below, in Sections~\ref{sec:highvalue} and \ref{sec:medvalue} we consider different priorities from users' and SPs' perspectives for authentication to high- and medium-value accounts (low-value accounts are less relevant to this discussion, since they do not require any of the more complex solutions as discussed herein). In Section~\ref{sec:detectimpersonation} we discuss the threat of user impersonation by rogue insiders at IdPs, and provide two examples to illustrate the complexity trade-offs associated with mitigating this threat.

\subsection{High Value SP Accounts} \label{sec:highvalue}

\textbf{User's perspective.} For high-value SPs, such as banking or government services, the threat of impersonation and/or data collection by IdPs (which relates to \emph{\notpimpersonation} and \emph{\privatefromidp}) is an important concern from users' perspectives. In other words, \emph{high value SP accounts require more protection against potential IdP misbehaviour}. This motivates the use of CM-based schemes, such as Firefox Sync or FIDO UAF, that manage user credentials on the user's own device.

\textbf{SP's perspective.} If relying on third-party IdPs, high value SPs may have concerns about IdPs following necessary security requirements (e.g., maximum number of authentication attempts before account lock-out, use of additional authentication factors), increasing the importance of \emph{\authidps} and \emph{\signalsloa}. However, \emph{\notpimpersonation} would be a concern not only for users as discussed above, but also for SPs. This leaves SPs with the following alternatives that they may pursue:
\begin{enumerate}
\item \textbf{Credential vault.} SPs can encourage users to follow good password management practices, including the use of a password manager, but this is difficult to enforce. Many high-value SPs (including many major banks) encourage their customers (and sometimes have made it mandatory \cite{krebs_trusteer}) to install client-side software such as IBM Trusteer Rapport (currently installed on several hundreds of thousands of systems~\cite{ibm_trusteer_2016}), to defend against client-side threats such as phishing or malware. However, numerous online comments by users indicate that such tools (which may also be extended to, e.g., verify the presence of a suitable password manager on the user's device) often degrade client-side performance and stability. FIDO-certified UAF authenticators present a new opportunity for such institutions to enforce client-side security requirements while being less intrusive and more usable for end-users.
\item \textbf{FIS with strong IdP vetting.} High-value SPs that prefer a FIS scheme to delegate authentication to third-party IdPs must coordinate extensively (e.g., by putting in place appropriate auditing mechanisms) to compensate for the lack of the \emph{\notpimpersonation} benefit. SecureKey Concierge is an example of such a scheme (with millions of users~\cite{skconcierge_numusers}), which has extensive technical requirements for IdPs, currently consisting of large Canadian financial institutions. Partnering with financial institutions as IdPs has the unique advantage of benefiting from their fraud detection mechanisms---compromise in banking credentials leading to fraudulent transactions is likely to be caught quickly (either by automated means, or by users who notice transactions they did not authorize) and would result in the bank issuing new credentials for users. This provides additional protection to SPs against long-term compromise of accounts.
\end{enumerate}

\subsection{Medium Value SP Accounts} \label{sec:medvalue}

\textbf{User's perspective.} For medium-value accounts, such as online shopping or blogging platforms, users may be more concerned about SP misbehaviour such as spamming or misuse of profile information obtained from IdPs \cite{sun2011usersrefuse}. In other words, \emph{more protection is required against SP misbehaviour} (as opposed to IdP misbehaviour; cf. Section~\ref{sec:highvalue}), thereby increasing the importance of the \emph{\spunlinkable} and \emph{\nodatasharing} benefits. Protection against IdP misbehaviour is likely a lower priority to users for medium-value SP accounts: the majority of such SPs already communicate extensively with users by e-mail (e.g., sending invoices and other detailed account information), and the vast majority of such SPs already use e-mail based password reset \cite{bonneau2010thicket}. Therefore, users already place extensive trust in their e-mail providers when managing their SP accounts. This motivates the use of e-mail providers as IdPs for medium-value SP accounts.

Deployment of FIS schemes that allow users to use their e-mail addresses as their online identity and that minimize sharing of user data are scarce, e.g., Mozilla Persona was not adopted by either IdPs or SPs. Moreover, even if such a scheme were to be widely deployed, it may eventually be superseded by a scheme that is less privacy-friendly, e.g., major OpenID 2.0 IdPs migrated \cite{google_openid_migrate} their users to OpenID Connect (which enables more extensive sharing of user profile information), thereby forcing users to either accept the change or to discontinue use of their existing SP accounts and create new SP accounts. Such scenarios also raise the issue of the lack of \emph{\portableid}, which anchors users to their IdPs; another example is if a user uses their mobile network operator (MNO) as their IdP (e.g., via Mobile Connect), it will be more cumbersome to switch their mobile plan to a different MNO, since the user will lose access to all their SP accounts. Considering all these drawbacks, we believe the optimal solution from a user's point of view is the same as for high-value accounts: CM-based schemes such as Firefox Sync or FIDO UAF.

\textbf{SP's perspective.} SPs have an incentive to push users towards FIS schemes, to reduce the effort required in the sign-up process (e.g., the user is not required to select a new password) and to gain access to users' social media circles. The latter benefit is especially appealing to smaller SPs that have limited resources to do their own collection of user data. Medium-value SPs are more likely to choose the schemes that maximize the size of their user base, and therefore many SPs that offer the option for FIS-based SSO also offer the option for users to create a ``conventional'' account by selecting a user name and password.

\subsection{Detecting Impersonation by IdP} \label{sec:detectimpersonation}
As discussed above in Section~\ref{sec:medvalue}, many SPs allow users to reset their passwords by e-mail. This theoretically would allow e-mail providers to impersonate users by performing a password reset---however, this would cause the user to lose access to their account, and is therefore a detectable impersonation attempt. Detectability of IdP impersonation may provide a sufficient deterrent from doing so. Deterrence as a tool to hold trusted third-parties accountable can be found in other contexts as well, such as certificate transparency~\cite{cert_transp} for TLS certificate authorities (CAs).\footnote{Certificate transparency requires TLS CAs to publicize every certificate that they digitally sign, enabling detection of maliciously-issued certificates.} An important distinction between the CA model and the FIS model is that CAs can generally issue certificates for \emph{any} domain, whereas IdPs can only impersonate their own users (but not users of other IdPs). However, since the vast majority of the FIS market share is held by a small number of IdPs, a rogue IdP can still have a major impact; this motivates the deployment of countermeasures, or at least deterrence or accountability measures, against impersonation. Below, we discuss how two of the schemes analyzed can be augmented with existing tools to offer detection of impersonation by IdPs. The trade-off associated with this benefit is that it requires SPs to maintain additional state information, a shared secret, or out-of-band channel (i.e., independent from the IdP) with the user.

\subsubsection{Augmenting SAW}
When users log into an SP with SAW, the SP e-mails an OTP to users. An administrator from within an e-mail provider could therefore impersonate any of their own users by initiating an authentication process to an SP and intercepting the OTP e-mailed from the SP to the user. A possible deterrent to this threat would be for the SP to use client-side cookies (and with TLS token binding \cite{ietf_tokenbinding} for added security against cookie theft) to recognize whether or not it is the first time that the user has logged in from a browser; if not, the SP can send an out-of-band notification to the user informing them (or requesting confirmation, for added security) that their account has been accessed from a new machine. The out-of-band notification could be sent via, e.g., a secondary e-mail address, SMS, a push notification directly to the user's browsers on their existing devices (via the W3C Push API~\cite{w3cpushapi}), or instant messaging to users' social media accounts.

\subsubsection{Augmenting Mobile Connect} \label{sec:mc_detectimpersonation}
One of the available LoA3 authentication options in Mobile Connect involves using the Mobile Signature Service (MSS)~\cite{etsi_tr102_203} standard to generate and store an asymmetric key pair on the user's SIM card, to be used for authenticating the user's device to the mobile network operator (MNO) \cite{mobileconnect_openid}. When users log into an SP via Mobile Connect, they are first redirected to their IdP (i.e., their MNO), which initiates an authentication process whereby a confirmation prompt is received on the user's mobile device, and the user's response (signed using the key stored on the SIM) is sent back to the IdP. The IdP then generates an access token and redirects the user back to their SP, completing the standard OpenID Connect protocol flow as discussed in Section~\ref{sec:sso_protocols}.

An alternative Mobile Connect design offering resistance to impersonation by IdPs could leverage the cryptographic capabilities of SIM cards (currently only used for \emph{device-to-IdP} authentication, as described above) to implement a \emph{device-to-SP} cryptographic challenge-response protocol. A possible implementation would be for the user device to generate unique site-specific cryptographic key pairs (or use a signature scheme that uses a single key as a seed to generate unique site-specific keys \cite{ibm_identity_mixer,camenish_signature_2002}), similarly to FIDO UAF. However, IdPs could sign users' public keys, to attest to SPs that users' signing keys are generated on an IdP-authenticated SIM. If a user loses their SIM, the IdP could issue the user a new SIM; this would require the user's device to generate new keys for each SP on their new SIM. A disadvantage of this approach is that it requires a mechanism for key revocation, whereby IdPs inform SPs that users have lost their devices so that SPs can invalidate users' old keys and accept the newly-generated keys. An advantage of this alternative approach is that the key revocation and regeneration process enables IdPs to facilitate account recovery in the event that users lose their mobile phones (in contrast to schemes such as FIDO UAF, in which IdPs cannot facilitate account recovery, and which thereby require SPs to implement their own recovery mechanisms). With this alternative design, user impersonation by rogue IdPs would require revocation of targeted users' SP-specific keys, which would be easily detectable by users (i.e., they would lose access to their accounts, similar to the scenario of a rogue e-mail provider performing a password reset).

\subsection{Configuration for Device-Based (T2) Schemes}
As discussed above in Section~\ref{sec:detectimpersonation}, countermeasures to IdP impersonation may require additional protocol and implementation complexity for IdPs and SPs. On the other hand, SSO is also achievable through user-to-device (T2) authentication mechanisms, which eliminates the requirement for users and SPs to trust remote IdPs for authentication (T1); however, this requires additional client-side configuration effort. As demonstrated in Table~\ref{table:sso_categorization}, no device-based (T2) CM scheme (A5, A6) fully offers \emph{\nodevicepair}.

Developing usable device configuration mechanisms can be a challenge. For example, a very high proportion of Firefox Sync 1.5 (see Section~\ref{sec:sync_overview}) users had only set up a single device---a practice which would result in those users losing all their passwords if they were to lose their device or had to re-install their operating system (assuming, as revealed by a high volume of user complaints \cite{warner_ffsync}, that most of those users had not backed up their password database encryption key). Moreover, many users did not understand the purpose of ``pairing'' (i.e., transferring the encryption key to) new devices. On the other hand, Firefox Sync 2.0 simplifies configuration by using a password-derived encryption key, which improves usability but introduces weakness against both online and offline guessing attacks.

FIDO UAF also requires configuration for new devices. However, since the cryptographic keys cannot be extracted and transferred to new devices, users must ``enroll'' their new device individually on each SP account. For example, Microsoft's SP developer guide \cite{windows_hello_dev} for application developers suggests that new devices can be enrolled by asking the user to authenticate via a password and a second factor such as an SMS or e-mailed OTP, after which the new device can generate a new cryptographic key pair and send the corresponding public key to the SP. While it is too early to tell with certainty what authentication mechanism the majority of SPs will use when enrolling new UAF devices, relying on password-based authentication would nullify most of the security and usability benefits of using UAF, since: (1) attackers would focus on breaking the password authentication option instead of breaking UAF authentication, and (2) users who already use a password manager to automatically fill in strong passwords would see no usability or security benefit in enrolling any UAF devices to their SP accounts.

\section{Concluding Remarks} \label{sec:conclusion}
As we expected, no scheme in our classification and analysis of 14 web SSO schemes emerges as a clear winner for all use cases. Given the major differences in requirements across different use cases (e.g., see comparison of medium-value and high-value accounts in Sections~\ref{sec:medvalue} and \ref{sec:highvalue}), this should not be surprising. However, for the foreseeable future, we believe that credential manager (CM-based) SSO schemes such as Firefox Sync 2.0 emerge as a well-rounded solution that offers users improved security over password authentication with minimal usability impact. 

While CM-based schemes may be the overall best option from a user's perspective, practical mechanisms are lacking for service providers (SPs) wishing to require the use of specific CM schemes by their users and to ensure the use of suitably-secure client systems. SPs that need to enforce higher security guarantees (and have the resources to do so) require other schemes; e.g., with sufficient co-ordination, large institutions such as governments, banks, and telecommunications companies can form partnerships to enable stronger authentication without imposing a usability burden on users, as illustrated by the success of a number of federated identity systems (FIS schemes) around the world such as SecureKey in Canada, GOV.UK Verify in the UK \cite{deloitte_blueprint,govuk_verify}, and BankID in Sweden and Norway \cite{bankid}. FIDO UAF~\cite{fido_uaf} is another option that allows SPs to enforce stronger user authentication (vs. passwords), but the necessity for SPs to rely on a backup authentication mechanism for enrolling new devices will, in practice, result in a wide variation of the overall security and usability benefits gained (based on the backup mechanism used). Therefore, UAF-like schemes only offer meaningful benefits in highly specific scenarios; e.g., SPs that can eliminate password authentication for enrolling new devices, and instead rely on more secure (but typically less usable) mechanisms, and also have the resources to offer the additional user support that such a decision will require (i.e., organizations that are better insulated from the consequences of poor usability \cite{florencio_policies_2010}).

Table~\ref{table:sso_categorization} highlighted trade-offs that must be made when designing an SSO scheme, based on the benefits being prioritized. For example, CM-based schemes (A5 and A6) typically offer \emph{\portableid}, \emph{\restempoutage}, and more of the privacy benefits (B12-B14); on the other hand, A1-A4 schemes are better suited when \emph{\nodevicepair} is required. Section~\ref{sec:mc_detectimpersonation} further highlighted that different stakeholders (namely users, IdPs, and SPs) may have conflicting priorities; for example, \emph{\notpimpersonation} is beneficial for users but more complex for IdPs and SPs to implement. We also wish to emphasize that protocols that offered ``niche'' benefits to users, such as OpenID 2.0 (offering the freedom to users to use any IdP at any SP) or Mozilla Persona (offering the ability to browse privately from IdPs) have not survived the market. Instead, OAuth based protocols have emerged as the most dominant; we believe that this is due in part to its higher suitability towards providing access to users' social media data (detrimental for users, but useful to SPs and IdPs). While Table~\ref{table:sso_categorization} summarizes our analysis of the 14 SSO schemes examined herein, we hope that the framework from which it is derived proves to be useful in general for evaluating new schemes in the future.

\section*{Acknowledgements}

This work was carried out while the first author was a PhD candidate at Carleton University, and revised while a faculty member at the University of Toronto Mississauga. The second author acknowledges funding from the Natural Sciences and Engineering Research Council of Canada (NSERC) for both his Canada Research Chair in Authentication and Computer Security, and a Discovery Grant. 

\bibliographystyle{ACM-Reference-Format}

\end{document}